\documentclass[11pt,a4paper]{article}

\usepackage[headheight=15pt,left=2cm,right=2cm,top=2cm,bottom=1.5cm]{geometry}
\usepackage{sectsty}

\usepackage{apacite}
\usepackage{amsmath}
\usepackage{amssymb,bm}
\usepackage{amsthm}
\usepackage{natbib}
\usepackage{multirow}
\usepackage{tabularx}
\usepackage{booktabs}
\usepackage{arydshln}
\usepackage{graphicx}
\usepackage{etoolbox}
\usepackage{placeins}
\usepackage{enumerate}
\usepackage{color}
\usepackage{fancyvrb}
\usepackage{bbm}
\usepackage{hyperref}
\usepackage[capitalise]{cleveref}
\usepackage{dcolumn}
\usepackage{mathabx}
\usepackage{comment}
\usepackage{soul}
\usepackage{setspace}
\usepackage{xcolor}
\usepackage{subcaption}
\usepackage{caption}
\usepackage{rotating}

\newcolumntype{R}{>{\raggedleft\arraybackslash}X}

\newcommand{\Cor}{\mathrm{Cor}}

\usepackage{authblk}
\usepackage{color,soul}

\usepackage{blindtext}

\usepackage{subfiles} % Best loaded last in the preamble

\title{Multilevel modelling of double-sampled clustered social networks with individual-level data on between-cluster ties}

\author{Fiona Steele\footnote{Corresponding author. Email: f.a.steele@lse.ac.uk} \\ \small{\it{Department of Statistics, London School of Economics \& Political Science, UK}}
\and Justin Weltz \\ \small{\it{Santa Fe Institute, USA}} \\
\and Eleanor A. Power \\ \small{\it{Department of Methodology, London School of Economics \& Political Science, UK; Santa Fe Institute, USA}} \\
\and Jeremy Koster \\ \small{\it{Department of Human Behavior, Ecology, and Culture, Max Planck Institute for Evolutionary Anthropology, Germany}} }
\date{}

\begin{document}
\maketitle

\begin{abstract}
We consider the analysis of dyadic network data on ties between individuals in different clusters where the presence of a directed tie is reported by each individual in a dyad and the cluster-level network is of interest. A generalisation of the Social Relations Model (SRM) is proposed which includes actor, partner and dyad effects at the individual and cluster levels. The model additionally uses ``double-sampling'' of ties to estimate a measurement model which adjusts for and quantifies the extent of reporter effects.  The model can be viewed as a type of multilevel structural equation model, with multiple cross-classified random effects, which can be estimated using Markov chain Monte Carlo (MCMC) methods in Bayesian software.  Using parameter estimates from this multilevel SRM, we then propose two alternative ways of deriving the between-cluster network that are based on predictions of the strength of between-cluster ties.  Our approach is illustrated using data on social support networks in a rural community in Nicaragua where individual reports of bidirectional exchanges of support with individuals from other households are used to derive the between-household network.
\end{abstract}

\section{Introduction}
\label{sec:intro}

Social network data are collected in a range of fields such as anthropology \citep[e.g.][]{ready.power2018}, economics \citep[e.g.][]{banerjee.etal2013}, management \citep[e.g.][]{tsai.1998} and sociology \citep[e.g.][]{entwisle.2007}, providing information on relationships between pairs of subjects, or ``dyads''. Data on dyadic relationships are often bidirectional (i.e. directed), leading to two relationships or ``ties'' for each dyad, one associated with each dyad member in their distinct roles as ``actor'' and ``partner''. In a support network, for example, a dyad formed of individuals A and B would have one tie representing support from A (the actor) to B (the partner) and a second for support from B to A. A common practice when gathering directed social network data through self-reports is to ``double sample'' relationships \citep[e.g.][]{reagans2003network,nolin2010food,ready.power2018}.  For individual dyad (A,B), for example, each of A and B report on both support from A to B and on support from B to A from their dual perspectives as actor and partner, yielding two reports of each directed relationship and four responses per dyad.
Meanwhile, as for many other forms of social data, network data may exhibit clustering, such as individuals nested in families, households, organisations, or geographical regions. While data are often available only on relationships between individuals in the same cluster, more generally dyads may be formed of individuals from different clusters, thus providing valuable information on the between-cluster network.  

The household is one such cluster that is of key interest to social scientists. As the unit to which resources are typically pooled, the household is often the fundamental object of study for investigations of wealth, economic mobility, intergenerational inheritance, etc. \citep{netting.etal1984}. Researchers with an interest in understanding household wealth dynamics, for instance \citep[e.g.][]{pfeffer.waitkus2021}, may be interested not only in measures of household production and consumption, assets, and demographic structure, but also of households' social position or social capital. The extent to which a household is able to call upon support, is embedded in dense networks of mutual exchange, or is able to serve as a broker within the community could all plausibly contribute to its ability to accumulate wealth, for example. As measures of assets are inherently at the level of the household, it follows that network measures should be at the equivalent level, reflecting the collective relationships and connections of household members. Household networks have also been studied in health research, for example to assess the impact of household-level social support networks on individuals' uptake of preventative chemotherapy in Uganda \citep{chami.etal2017} and incidence of diarrhoeal disease in Ecuador \citep{bates.etal2007}. Other examples of cluster-level networks based on individual-level data arise in organisational research where there is interest in the study of knowledge exchange both within and between organisations. In a review of inter-organisational networks, \cite{sorenson2014organizations} argue that relationships between organisations are driven by informal interpersonal connections between employees from different organisations where there may be multiple individual points of contact between organisations. There may also be interest in the relationships between units (i.e. clusters of employees) within an organisation \citep[e.g.][]{hansen1999search}.

Our research is motivated by the study of support networks in the social sciences where individuals are clustered in households and data are collected from individuals on support given to and received from individuals in other households. An individual reports on exchanges of support with other individuals from their perspective as both the giver (actor) and receiver (partner), leading to double-sampling of directed ties. We focus on two methodological questions: 1) How can double-sampled data be used to study and adjust for reporter effects in analysis of the between-individual network? 2) How can data collected at the individual level be used to derive estimates of the between-household network? Previous research has considered these questions separately using simple aggregation methods.  In the case of multiple reports from double-sampling, the typical approach is either to take the union (allowing ``unilateral nominations'') or intersection (requiring ``mutual assent'') \citep{krackhardt.1987, lee.butts2018}. Simplistic methods of network aggregation, however, can influence the resulting structure of the network, due not only to the changes in the resulting number of edges, but also due to respondents' tendency to name the same people across the distinct questions; for example, taking either the union or intersection can putatively result in substantially higher levels of reciprocity \citep{ready.power2021}. Consequently, a number of models exist that use such multiple reports to directly model the uncertainty in the existence of the ``true'' latent network \citep{an.schramski2015, lee.butts2018, debacco.etal2023, redhead.etal2023}. These, however, do not account for the nested nature of responses, when individuals are situated in households (or other clusters). Turning to the second question, the standard approach to constructing a household-level network from individual-level data is a variant of the union method used for aggregating double-sample data to the individual level: a binary response is defined for each pair of households indicating whether there is a tie between any individual in one of the households and any individual in the other \citep[e.g.][]{banerjee.etal2024}, where ties may be directed or undirected.  This approach is wasteful, however, as it does not make full use of the individual-level information.  Moreover, we show that simple aggregation may lead to biased estimates of the probabilities of between-household ties if data are collected on a sample of eligible individuals in surveyed households.  

In this paper, we propose a multilevel model for double-sampled individual-level dyadic binary data on between-cluster ties, taking the household as our exemplar cluster. Our model is a type of multilevel structural equation model with measurement and structural model components.  The measurement model treats the double-sampled reports of each directed between-individual tie as two noisy measurements of the latent propensity of a ``true'' directed tie and includes individual- and household-level reporter effects. The structural model specifies this latent tie propensity as a function of individual- and household-level actor, partner and dyad effects, in a multilevel extension of the well-known Social Relations Model (SRM) \citep{kenny.lavoie1984}. The proposed model is a complex non-hierarchical multilevel model with multiple cross-classified random effects that can be estimated using Markov chain Monte Carlo (MCMC) methods in Bayesian software.  We describe how predicted probabilities of individual ties can be computed from the fitted model and used to derive measures of the strength of directed between-household ties (edge weights) which define a household network.  We consider two alternative between-household edge weights which are model-based extensions of simple descriptive methods of aggregation, and illustrate their application in the construction and analysis of a household social support network in a rural Nicaraguan community.

Section 2 describes the structure of clustered dyadic data on double-sampled directed ties and introduces notation.  The multilevel SRM is set out in Section 3, and alternative approaches to building household networks are presented in Section 4. The application of the proposed methods is illustrated in Section 5 using data on social support ties between individuals from different households.  A concluding discussion, including potential directions for future research and recommendations for collecting data on network ties, is given in Section 6. 

\section{Data structure and notation}
\label{sec:data_structure}

Before setting out methods for the analysis of double-sampled clustered social network data, we describe the data structure and introduce the notation used in subsequent sections. For ease of exposition, we describe the data and the statistical model in terms of our application to social support networks where the presence of a tie between individuals represents an exchange of support and $n$ individuals are clustered in $H$ households.

We consider the case of bidirectional exchanges where the individual giving help is referred to as the ``actor'' and the individual receiving as the ``partner'' (other common labels given to these roles include ``sender'' and ``receiver'').  Moreover, each member of a dyad reports on exchanges with the other from their perspective as both actor and partner, leading to ``double-sampled'' data (or dual reports). 

Denote by $y_{i,j,m}$ a binary response coded 1 if individual $i$ (the actor) is reported to give help to individual $j$ (the partner) by reporter $m \in \{i,j\}$. For dual reports of bidirectional exchanges, we have four responses per dyad $(y_{i,j,i}, y_{i,j,j}, y_{j,i,j}, y_{j,i,i})$ for $i,j = 1, 2, \ldots, n$ and $i < j$.
Denote by $h(i)$ and $h(j)$ the households of individuals $i$ and $j$, referred to as the actor and partner households for the individual dyad $(i,j)$.  We will also index the actor and partner households by $k$ and $l$ respectively for $k, l = 1, 2, \ldots, H$ when we refer to a household dyad without reference to their individual members.  Let $n_k$ and $n_l$ denote the number of individuals sampled from households $k$ and $l$.

To illustrate the data structure, Table \ref{tab:data_structure} shows the records for help given by three individuals ($i=1,2,3$) in household $k=1$ to two individuals ($j=4,5$) in household $l=2$, where each of the six ties is reported from the perspective of the actor ($m=i$) in household 1 and the partner ($m=j$) in household 2, leading to $2 n_k n_l = 12$ records.  The table does not show the additional 12 records corresponding to the help given by individuals in household 2 to individuals in household 1, although they are also included in the analysis sample. In the case of undirected ties, household dyads $(k,l)$ and $(l,k)$ are exchangeable, and $y_{i,j,m} = y_{j,i,m}$, in which case the individual exchanges between households $k$ and $l$ can be compressed to $n_k n_l$ records.  This is an example of the ``long'' form data structure.  Alternatively, the data can be analysed in ``wide'' form with one record per directed dyad, and the actor and partner reports stored in separate variables.

\begin{table} [!htbp]
\caption{Example of ``long'' data structure showing dual reports $y_{i,j,m}$ of presence of directed tie from individual $i$ in household $k$ to individual $j$ in household $l$ where $n_k=3$ and $n_l=2$. The records for ties from $j$ to $i$ are suppressed for brevity.}
\centering
\label{tab:data_structure}
\begin{footnotesize}
\begin{tabular}{cccccc}
\hline
Actor $i$ & Partner $j$ & Reporter $m \in (i,j)$ & Actor hh $h(i)=k$ & Partner hh $h(j)=l$ & $y_{i,j,m}$ \\
\hline
1 & 4 & 1 & 1 & 2 & 1 \\
1 & 4 & 4 & 1 & 2 & 1 \\
1 & 5 & 1 & 1 & 2 & 0 \\
1 & 5 & 5 & 1 & 2 & 0 \\
2 & 4 & 2 & 1 & 2 & 1 \\
2 & 4 & 4 & 1 & 2 & 0 \\
2 & 5 & 2 & 1 & 2 & 0 \\
2 & 5 & 5 & 1 & 2 & 1 \\
3 & 4 & 3 & 1 & 2 & 0 \\
3 & 4 & 4 & 1 & 2 & 1 \\
3 & 5 & 3 & 1 & 2 & 0 \\
3 & 5 & 5 & 1 & 2 & 1 \\
\hline
\end{tabular}
\end{footnotesize}
\end{table}

We focus on between-household ties where $k \neq l$, but later comment on how to incorporate information on within-household ties.
We also restrict attention to binary ties, though the proposed methods can be generalised to valued ties such as counts (e.g. the number of times that $i$ helped $j$ in the last six months) or continuous measures of tie strength (e.g. the amount of money that $i$ gave to $j$).

The full dataset with the individual dyads for all $H(H-1)$ directed household dyads has a cross-classified structure: the $n$ individuals are members of multiple individual dyads and therefore multiple household dyads.  For example, individual 1 in Table 1 is a member of two directed individual dyads as actor when paired with the members of household 2, and a further $\sum_{h=3}^{H} n_h$ individual dyads when paired with the members of the remaining $H-2$ households, and for each of these dyads there are two reports of the presence of a tie from $i$ to $j$.

\section{Multilevel Social Relations Model for double-sampled clustered network data}
\label{sec:multilevel_SRM}

We specify a model that accounts for double-sampling of directed ties between individuals, nesting of individuals within households, and the cross-household nature of individual ties. Denote by $y_{i,j,m}^*$ a continuous latent variable underlying the observed binary $y_{i,j,m}$ such that $y_{i,j,m}=\mbox{I}(y_{i,j,m}^*>0)$ where $\mbox{I}(\cdot)$ is the indicator function. We assume that ties are directed and specify a multilevel structural equation model with two components: a measurement model for the relationship between $y_{i,j,m}^*$ and a latent variable $z_{i,j}$ representing the underlying propensity of a directed tie from individual $i$ to individual $j$ (e.g. that $i$ helps $j$), denoted by $i \rightarrow j$, and a structural model for $z_{i,j}$ which takes the form of a multilevel Social Relations Model.  In both model components, an individual- and household-level effect is denoted by superscript (1) and (2) respectively and, as in Section \ref{sec:data_structure}, $h(i)$ and $h(j)$ denote the households of individuals $i$ and $j$.  For ease of reference, a table summarising key notation is given in the Appendix. The required data structure for estimation, including the form of the identifiers for all random effects, is given in the supplementary material (Section S1).

\subsection{Measurement model}
\label{subsec:multilevel_SRM_measurement}
We specify a probit measurement model for $y_{i,j,m}^*$, using dual reports of ties to quantify and adjust for reporter effects:
\begin{eqnarray}
\label{eq:model_measurement}
y_{i,j,m}^* & = & z_{i,j} + \mbox{I}(m=i) (\beta_a + r_{ai}^{(1)} + r_{a,h(i)}^{(2)}) + \mbox{I}(m=j) (\beta_p + r_{pj}^{(1)} + r_{p,h(j)}^{(2)}) \\ \nonumber
& + & r_{d |i,j| m}^{(1)} + r_{d |h(i),h(j)| h(m)}^{(2)} + \epsilon_{i,j,m}, \quad
i,j=1,\ldots,n; m \in \{ i,j \}.
\end{eqnarray}
This is a generalisation of the popular Rasch model \citep{rasch1980} that allows for individual- and household-level actor, partner and dyad-specific errors in the reporting of ties.  The observed responses $(y_{i,j,i},y_{i,j,j})$ from the two reporters are treated as measurements of a latent variable $z_{i,j}$ which represents the true propensity that individual $i$ helps individual $j$, with intercepts $(\beta_a,\beta_p)$ that allow the overall probability of reporting a tie to depend on whether the reporter $m$ is the actor or partner.  We distinguish individual- and household-level components of measurement/reporting error according to whether the reporter is the actor or partner:  $r_{ai}^{(1)}$ is the tendency of individual $i$ to report giving help to others (i.e. when reporting as the actor), $r_{a,h(i)}^{(2)}$ is the tendency of individuals in $i$'s household to report giving help, while $r_{pj}^{(1)}$ and $r_{p,h(j)}^{(2)}$ are the corresponding individual- and household-level tendencies to report receiving help (i.e. when reporting as the partner).  We additionally allow for dyadic reporter effects at the individual and household level: $r_{d |i,j| m}^{(1)}$ captures correlation between reports by individual $m$ of exchanges of support (in either direction) with a specific person (i.e. between their reports of $i \rightarrow j$ and $j \rightarrow i$ ties), while $r_{d |h(i),h(j)| h(m)}^{(2)}$ captures correlation between reports by members of $m$'s household of their exchanges with members of a specific household.  Finally, $\epsilon_{i,j,m}$ is a residual specific to $m$'s report of directed tie $i \rightarrow j$.

We make the following distributional assumptions:
\begin{equation}
\label{eq:measurement_dist}
\begin{gathered}
\begin{pmatrix}
r_{ai}^{(1)} \\
r_{pi}^{(1)} \\
\end{pmatrix}
\sim N \left ( \bm{0},
\Sigma_{rap(1)}
\right ), \quad
\Sigma_{rap(1)} = \begin{pmatrix}
\sigma_{ra(1)}^2 & \\
\sigma_{rap(1)} & \sigma_{rp(1)}^2 \\
\end{pmatrix}\\[2ex]
\begin{pmatrix}
r_{a,h(i)}^{(2)} \\
r_{p,h(i)}^{(2)} \\
\end{pmatrix}
\sim N \left ( \bm{0},
\Sigma_{rap(2)}
\right ), \quad
\Sigma_{rap(2)} = \begin{pmatrix}
\sigma_{ra(2)}^2 & \\
\sigma_{rap(2)} & \sigma_{rp(2)}^2 \\
\end{pmatrix}\\[2ex]
r_{d |i,j| m}^{(1)} \sim N(0, \sigma_{rd(1)}^2), \quad 
r_{d |h(i),(j)| h(m)}^{(2)} \sim N(0, \sigma_{rd(2)}^2), \quad\epsilon_{i,j,m} \sim N(0,1).
\end{gathered}
\end{equation}

For both $r_{ai}^{(1)}$ and $r_{pi}^{(1)}$ a positive value corresponds to an above-average tendency to overstate a tie as actor and partner respectively, while a negative value corresponds to an above-average tendency to understate a tie.  The covariance $\sigma_{rap(1)}$ allows for a correlation between an individual's reporting tendencies as actor and partner; for example, $\sigma_{rap(1)} < 0$ if individuals who tend to overstate help given also tend to understate help received. Similarly, at the household level, the covariance $\sigma_{rap(2)}$ allows for correlation between a household's actor and partner reporting tendencies, based on the reports of individuals in the same household.

Conditional on $z_{i,j}$, (\ref{eq:model_measurement}) allows for the following sources of correlation among the latent responses underlying reports of ties for the same dyad: correlation between $ y_{i,j,m}^*$ and $y_{j,i,m}^*$ due to correlation between $m$'s reporting tendencies as actor and partner and to $m$'s tendency to report in a similar way on the existence of ties $i \rightarrow j$ and $j \rightarrow i$, and correlation between the actor's and partner's reports of the same directed tie. Conditional on $z_{i,j}$ and $z_{i,j^\prime}$, the model also allows for correlation between $y_{i,j,i}^*$ and $y_{i,j^\prime, i}^*$ for $j \neq j^\prime$ due to $i$'s reporter effect as actor.  Similarly, conditional on $z_{i,j}$ and $z_{i^\prime,j}$, we allow for correlation between $y_{i,j,j}^*$ and $y_{i^\prime,j,j}^*$ for $i \neq i^\prime$ due to $j$'s reporter effect as partner. More generally, the model allows for a non-zero correlation between any pair of responses $y_{i,j,m}$ and $y_{i^\prime,j^\prime,m^\prime}$ that have at least one individual, or members of the same household, in common. The full covariance structure implied by the model is given in supplementary materials (Section S2). Finally, we make the standard conditional independence assumption of Rasch and other latent variable models that the correlation between $y_{i,j,i}^*$ and $y_{i,j,j}^*$ can be explained by their common dependence on $z_{i,j}$.

\subsection{Structural model}
\label{subsec:multilevel_SRM_structural}

The structural model for the (directed) latent tie propensity $z_{i,j}$ has the form of a multilevel Social Relations Model (SRM). The SRM was originally proposed by David Kenny and colleagues for the analysis of ``round robin'' dyadic data in psychology \citep{kenny.lavoie1984}, and has since found applications in a broad range of fields. We specify an extension to the traditional SRM with actor, partner and dyad effects at the individual and households levels. The model for the propensity of an $i \rightarrow j$ tie is specified as
\begin{equation}
\label{eq:model_structural}
z_{i,j} = a_i^{(1)} + p_j^{(1)} + d_{i,j}^{(1)} + a_{h(i)}^{(2)} + p_{h(j)}^{(2)} + d_{h(i),h(j)}^{(2)}
\end{equation}
where $a_i^{(1)}$ and $p_j^{(1)}$ are individual-specific actor and partner effects (shared by all dyads with the same actor or partner respectively), $d_{i,j}^{(1)}$ is an individual-level dyad effect specific to tie $i \rightarrow j$, $a_{h(i)}^{(2)}$ and $p_{h(j)}^{(2)}$ are household-specific actor and partner effects, and $d_{h(i),h(j)}^{(2)}$ is a household-level dyad effect specific to the between-household tie $h(i) \rightarrow h(j)$.  The model can be extended to include covariates, which may be characteristics of individuals or households (relating to the actor, partner or dyad).

We assume
\begin{equation}
\label{eq:structural_dist}
\begin{gathered}
\begin{pmatrix}
a_i^{(1)} \\
p_i^{(1)} \\
\end{pmatrix}
\sim N \left ( \bm{0}, \Sigma_{ap(1)} \right ), \quad
\Sigma_{ap(1)} = \begin{pmatrix}
\sigma_{a(1)}^2 & \\
\sigma_{ap{(1)}} & \sigma_{p(1)}^2 \\
\end{pmatrix} \\[2ex]
\begin{pmatrix}
d_{i,j}^{(1)} \\
d_{j,i}^{(1)} \\
\end{pmatrix}
\sim N \left ( \bm{0}, \Sigma_{d(1)} \right ), \quad
\Sigma_{d(1)} = \begin{pmatrix}
\sigma_{d(1)}^2 & \\
\sigma_{dd{(1)}} & \sigma_{d(1)}^2 \\
\end{pmatrix} \\[2ex]
\begin{pmatrix}
a_{h(i)}^{(2)} \\
p_{h(i)}^{(2)} \\
\end{pmatrix}
\sim N \left ( \bm{0}, \Sigma_{ap(2)} \right ), \quad
\Sigma_{ap(2)} = \begin{pmatrix}
\sigma_{a(2)}^2 & \\
\sigma_{ap{(2)}} & \sigma_{p(2)}^2 \\
\end{pmatrix} \\[2ex]
\begin{pmatrix}
d_{h(i), h(j)}^{(2)} \\
d_{h(j), h(i)}^{(2)} \\
\end{pmatrix}
\sim N \left ( \bm{0}, \Sigma_{d(2)} \right ), \quad
\Sigma_{d(2)} = \begin{pmatrix}
\sigma_{d(2)}^2 & \\
\sigma_{dd{(2)}} & \sigma_{d(2)}^2 \\
\end{pmatrix}.
\end{gathered}
\end{equation}

Correlations between the actor and partner effects at each level, denoted by $\rho_{ap{(1)}}$ and $\rho_{ap{(2)}}$, can be obtained in the usual way from the variances and covariances in $\Sigma_{ap(1)}$ and $\Sigma_{ap(2)}$; these are referred to as individual- and household-level generalised reciprocity correlations and both are expected to be positive, indicating that individuals (households) with an above-average tendency to give help tend also to have an above-average tendency to receive help.   The correlations between the individual-level dyad effects, $d_{i,j}^{(1)}$ and $d_{j,i}^{(1)}$, and between the household dyad-level effects, $d_{h(i), h(j)}^{(2)}$ and $d_{h(j), h(i)}^{(2)}$, are denoted by $\rho_{dd{(1)}}$ and $\rho_{dd{(2)}}$ and obtained from the elements of $\Sigma_{d(1)}$ and $\Sigma_{d(2)}$ respectively.  These correlations are referred to as dyadic reciprocity correlations and measure the extent to which an individual (household) that gives help to another individual (household) tends also to receive help from that individual (household); both are expected to be positive.

In a traditional multilevel SRM, ties are restricted to individuals in the same cluster, for example between members of the same family, leading to individual dyads nested within clusters \citep{snijders.kenny1999}. For such designs, the cluster (household) component of (\ref{eq:model_structural}) is reduced to a single random effect at the cluster level because cluster-level actor, partner and dyadic effects can be identified only when there are ties between individuals from different clusters.  \cite{koster2018} proposed an extension to the model of \citeauthor{snijders.kenny1999} that is suitable for single reports of between-cluster individual ties.  This model has a similar form to (\ref{eq:model_structural}), but with $z_{i,j}$ replaced by $y_{i,j,i}^*$ or $y_{i,j,j}^*$ and the addition of cross-level interaction random effects to allow for the possibility that an individual may be more likely to interact with some households than with others. We simplify the structural model to exclude cross-level effects, but extend Koster's model to include the measurement model for dual reports (\ref{eq:model_measurement}) to allow for reporter effects. (In principle, the measurement model could also include cross-level effects, which would reflect tendencies for respondents to nominate individuals from other households in clustered ways that are distinct from the other reporters in the respondents' households.) Our SEM approach is related to the dyadic Item Response Theory (dIRT) model which combines a measurement model for $z_{i,j}$ measured by multivariate ordinal responses and an SRM for $z_{i,j}$ \citep{gin.etal2020}. \citeauthor{gin.etal2020} also discuss briefly an extension of the dIRT that includes cluster-level actor and partner random effects, where individual ties can be within or between clusters, though an application is not presented.

\subsection{Models for undirected ties and within-household ties}
\label{subsec:multilevel_SRM_other_models}

The model described above is suitable for the analysis of directed ties.  For undirected ties, where $y_{i,j,m} = y_{j,i,m}$, simplifications of (\ref{eq:model_measurement}) and (\ref{eq:model_structural}) are necessary because there is no longer an actor-partner distinction: thus, in (\ref{eq:model_measurement}) $\beta_a=\beta_p$ and $r_{ai} + r_{pj}=r_i + r_j$, and in (\ref{eq:model_structural}) we can write $a_i^{(1)} + p_j^{(1)} = v_i^{(1)} + v_j^{(1)}$, $a_{h(i)}^{(2)} + p_{h(j)}^{(2)} = v_{h(i)}^{(2)} + v_{h(j)}^{(2)}$, $d_{i,j}^{(1)} = d_{j,i}^{(1)}$ and $d_{h(i),h(j)}^{(2)} = d_{h(j),h(i)}^{(2)}$.

The structural model of (\ref{eq:model_structural}) for between-household ties can also be generalised to incorporate ties between individuals in the same household. A joint model for between- and within-household ties is described in the supplementary material (Section S4).  This model was considered in preliminary analysis of Section \ref{sec:application} but, for our data, the predicted probabilities of within-household exchanges were close to 1, leading to wide credible intervals for parameters of the within-household component of the model. 

\subsection{Model identification and estimation}
\label{subsec:multilevel_SRM_estimation}
\subsubsection*{Identification}
Identification of the multilevel SRM is possible due to the round-robin study design with double-sampling of each directed dyad and clustering of individuals in households (see Table \ref{tab:data_structure} and Section S1 of the supplementary materials).  For example, the variances and covariances of the individual-level actor and partner reporter effects in (\ref{eq:model_measurement}) are identified by each individual reporting on their roles as actor and partner in multiple dyads, individual-level actor and partner effects in (\ref{eq:model_structural}) are identified by individuals' membership of multiple dyads, and households with more than one adult member contribute to estimation of the parameters of the corresponding household-level actor and partner effects. More formally, the model is identified if the random effect parameters can be uniquely identified by the variances and covariances of $y_{i,j,m}^*$ implied by the model (see \cite{gin.etal2020} for a discussion of identification for the related dyadic IRT model). Identification of the random effect parameters of the multilevel SRM is shown in Section S3 of the supplementary materials.  We also carry out a small-scale simulation study to assess whether the parameters of interest are reliably estimated from data generated to have the same multilevel structure as the Nicaraguan data analysed in Section \ref{sec:application}.  We find that we are able to recover the true values with accurate SEs for most parameters when the model assumptions hold. However, there is notable downward bias in the household-level actor-partner correlations $\rho_{rap(2)}$ (in the measurement model) and $\rho_{ap(2)}$ (in the structural model), which is likely explained by the relatively small number of households ($n=32$).  We therefore interpret these parameters with caution in the data analysis of Section 5.3.  Further details, including the simulation study results, are given in Section S3.

\subsubsection*{Estimation}
\cite{snijders.kenny1999} noted that the traditional SRM, with individual-level actor and partner effects, can be viewed as a type of cross-classified multilevel model where actor and partner effects are crossed because each actor is paired with multiple partners, and vice versa.  In our extended SRM, the household-level actor and partner effects are also crossed, and the structural model of (\ref{eq:model_structural}) must be jointly estimated with the measurement model of (\ref{eq:model_measurement}).  Bayesian MCMC methods provide a computationally efficient and flexible means of estimating cross-classified and other complex random effects models \citep{browne.etal2001} and are widely used for estimation of random effects models for dyadic data \citep[e.g.][]{hoff2005, koster2020statistical}. Another advantage of MCMC estimation is the possibility to derive estimates and credible intervals for functions of the parameters of the fitted model using the parameter chains, a feature used to compute variance partitioning coefficients in Section \ref{subsec:application_model_results} and predicted probabilities of between-household ties in Section \ref{subsec:deriving_cluster_network_SRM}.

All data analysis presented in Section \ref{sec:application} was carried out using Hamiltonian Monte Carlo sampling, as implemented in the Stan language \citep{carpenter2017stan} via the \texttt{rstan} R package \citep{guo2020package}.  The following prior distributions were specified: Normal priors with mean $0$ and precision (inverse variance) $0.0001$ for the actor report intercept and partner-actor report difference $(\beta_a, \beta_p-\beta_a)$, exponential priors with mean $1$ for the standard deviations of the univariate and bivariate random effects, and LKJ priors \citep{lewandowski2009generating} with shape $2$ for the correlation matrices of the bivariate random effects. For all random effects, we sample from standard normal distributions to improve mixing. Further details are given in Section \ref{subsec:application_model_results}.

\section{Deriving the cluster-level network}
\label{sec:deriving_cluster_network}

As noted in Section \ref{sec:intro}, a household-to-household network is often of direct analytical interest. Many development interventions, for example, are made at the household level \citep[e.g. an asset transfer to impoverished households as in][]{balboni.etal2022}, so tracking consequent impacts to social relations, or understanding how existing social relations might impact efficacy, require understanding household-level network connections. In such development studies, it is common for only one member of a household to be surveyed on behalf of the unit, due in part to logistical constraints \citep[e.g.][]{krishnan.sciubba2009, dexelle.holvoet2011, alatas.etal2012, jaimovich.2015, comola.prina2021, dexelle.verschoor2023}. However, such reliance on a single reporter may result in biases and myopia, as the single respondent may privilege or better recall their own relationships over those of other household members \citep{perkins2015social, simpson2022structural}. Some projects have therefore sought to expand the set of respondents from the household \citep[e.g.][]{banerjee.etal2013}. 
This could entail, for example, sampling both male and female household heads, or sampling all (or many) adult members. 
This more extensive information has the potential to produce more comprehensive networks, but raises new methodological questions about the operationalisation of network ties among households.

\subsection{Aggregation of individual-level data} 
\label{subsec:deriving_cluster_network_aggregation}

There are multiple methods for building a household network from individual-level data, 
and the choice between these constructions depends on the intended use of the network.
We consider networks as inextricably linked to dynamic processes that evolve over the nodes and edges they represent, e.g., a disease or an intervention. In this context, household membership provides important structure that can influence how phenomena spread through social connections \citep{aguiar2025illusionhouseholdsentitiessocial, koster2018}. We will consider a household ``infected" if every member of the household is influenced by the dynamic process of interest.
For example, imagine an intervention in which food items are given to random households in a Nicaraguan community. If these goods are shared by members of the household, this intervention immediately infects the sampled households. Additionally, researchers determine that these goods are likely to be shared along social support ties (and to impact all members of the households they are shared with). Consequently, the spread of this intervention across the individual-level network is dictated by household structure through patterns of shared resources.
For example, a household that receives access to formal financial institutions via a development intervention may subsequently be able to lend more to others \citep[e.g.][]{kinnan2012kinship} or not need to borrow as much money from others \citep{banerjee.etal2024}, indirectly impacting the finances (and relationships) of their social partners.
Therefore, the effect of household group membership on the probability of ``infection" 
depends on the nature of the spreading process. 

We define two representations of ties between households. The ``aggregation" approach constructs edge weights from the raw count of individual connections between households, modeling the overall connectivity between household members.
The ``density" approach builds edge weights from the number of individual connections relative to the total number of connections that could exist between households, 
capturing the general propensity for connections between household members. 
These methods could also correspond to dynamic processes with different household ``diffusion" dynamics 
\citep{aguiar2025illusionhouseholdsentitiessocial}; i.e. whether one infected individual implies everyone in their household is infected. 
If there is internal diffusion, then a household can be infected through any tie with an infected household. In this case, the aggregation method captures the appropriate connection strength for modelling network dynamics.
If there is no internal diffusion, then the interaction between an infected and non-infected household should be evaluated as the likelihood of the phenomenon spreading from a random individual in one household to a random individual in another household. In this context, the density approach is appropriate because the edge weights capture the probability of randomly selecting an individual from each household that are connected.

Previous research has primarily focused on a simplification of the aggregation method, defining a binary indicator $y_{k,l}^{u}$ of the presence of any tie between household $k$ and household $l$, coded 1 if \emph{any} individual in $k$ has a tie with \emph{any} individual in $l$ and 0 otherwise \citep[e.g.][]{banerjee.etal2024,bates.etal2007,chami.etal2017,hackman.kramer2021}. We refer to this approach as the ``union'' method. The union method is very sensitive to sampling structure. Suppose for simplicity that we observe one report (from either the actor or partner perspective) of the presence of a tie between individuals $i$ and $j$, denoted by $y_{i,j}$, where ties can be directed or undirected. Denote by $S_k$ and $S_l$ the set of sampled individuals from households $k$ and $l$, of size $n_k$ and $n_l$ respectively, then
\begin{equation}
\label{eq:y_kl_ag_def}
y_{k,l}^{u} = \mbox{I} \left (\sum_{i \in S_k, j \in S_l} y_{i,j} \geq 1 \right )
\end{equation}
where $\mbox{I}(\cdot)$ is the indicator function and $i<j$ for directed ties.
Let $y_{i,j}$ be a realisation of the random variable $Y_{i,j} \sim \mbox{Bernoulli}(\pi_{i,j})$ and $y_{k,l}^{u}$ a realisation of $Y_{k,l}^{u} \sim \mbox{Bernoulli}(\pi_{k,l}^{u})$.  From (\ref{eq:y_kl_ag_def}) the probability of a household tie based on aggregation is
\begin{equation}
\label{eq:pi_kl_sample}
\pi_{k,l}^{u} = \mathbb{P}(Y_{k,l}^{u}=1) = 1 - \mathbb{P} \left ( \sum_{i \in S_k, j \in S_l} Y_{i,j} =0 \right )
= 1 - \prod_{i \in S_k, j \in S_l} (1 - \pi_{i,j}).
\end{equation}
We can clearly see the dependence of $\pi_{k,l}^{u}$ on sample size by considering the special case where the individual tie probabilities are constant, i.e. $\pi_{i,j} = \pi$.  Then (\ref{eq:pi_kl_sample}) simplifies to
\begin{equation}
\label{eq:pi_kl_sample_constant}
\pi_{k,l}^{u} = 1 - (1 - \pi)^{n_k n_l},
\end{equation}
which increases with $n_k n_l$.

Our density and aggregation methods contain more information concerning the strength of connection between households than the ``any tie" household network. 
In our aggregation approach, we represent the edge weight between households $k$ and $l$ as $y_{k,l}^{\mbox{ag}} = \sum_{i \in S_k, j \in S_l} y_{i,j}$. 
In contrast, the density approach corresponds to the sample proportion of ties between households $k$ and $l$:
\begin{equation}
\label{eq:p_kl}
p_{k,l} = \frac{y_{k,l}^{\mbox{ag}}}{n_k n_l}.
\end{equation}
 If a full community census or a random sample is taken, then both edge weights are feasible. However, estimation of the household network becomes more complex when a household stratified sample is conducted. For example, if a random subset of fixed size is taken from each household, then the aggregation approach can no longer capture the size of the household (where household size is often understood to be the number of adults in the household). 
Alternatively, if the size of the random subset is proportional to the size of the household, then both edge weights can be estimated.

Regardless of the sampling scheme, both the density and aggregation constructions are limited by measurement error due to reporter effects. 
For example, the household dyad illustrated in Table~\ref{tab:data_structure} has an aggregation weight of $y_{k,l}^{\mbox{ag}} = 1$ if we consider only ties that both individuals in the dyad agree on and a weight of $y_{k,l}^{\mbox{ag}} = 6$ if all ties are used. Additionally, $y_{k,l}^{\mbox{ag}} = 2$ and $p_{k,l} = 1/3$ based on reports by actors, and $y_{k,l}^{\mbox{ag}} = 4$ and $p_{k,l} = 2/3$ based on reports by partners. More generally, measures of a household's centrality in the between-household network will be overstated (understated) if the household contains individual(s) with a tendency to over (under) report their ties with individuals in other households.
Consequently, we use a measurement model to account for this bias in self-reported connections and apply model-based density and aggregation approaches in Section~\ref{subsec:deriving_cluster_network_SRM}. 
Lastly, we note that $y_{k,l}^{\mbox{ag}}$ will generally be proportional to the sizes of households $k$ and $l$. This is logical under diffusion since infected households with more members have a higher likelihood of infecting other households.

\subsection{Deriving the between-household network from the multilevel SRM}
\label{subsec:deriving_cluster_network_SRM}

We propose a model-based approach to obtain household-level networks that extends the density and aggregation methods defined in Section~\ref{subsec:deriving_cluster_network_aggregation}.
This will be a principled method of transforming individual-level data into a household network while overcoming the measurement issues faced by $y_{k,l}^{\mbox{ag}}$ (and its binary form $y_{k,l}^u$) and $p_{k,l}$.
From the fitted multilevel network model described in Section~\ref{sec:multilevel_SRM}, we construct between-household networks using predicted probabilities of individual ties that account for reporter effects and thus measure the strength of ``true'' between-individual ties.
These model predictions are based on parameter estimates that are precision-weighted to allow for uncertainty when the number of observations is small.
The derived measures of the strength of between-household ties (edge weights) can then be used as inputs in the computation of summary statistics describing the household network (see Section \ref{subsec:application_network_analysis}).
Moreover, the MCMC chains for the model parameters and random effects can be used to obtain measures of uncertainty for predictions of network statistics (e.g. posterior standard deviations or credible intervals).  

We begin by deriving an expression for the marginal probability of a between-individual tie where the reporter effects are integrated out.  This marginal probability is obtained by partitioning the linear predictor of the multilevel SRM of (\ref{eq:model_measurement}) and (\ref{eq:model_structural}) as
\begin{equation*}
y_{i,j,m}^* = \eta_{i,j} + w_{i,j,m}
\end{equation*}
where
\begin{equation}
\label{eq:eta_ij}
\eta_{i,j}= 0.5(\beta_a + \beta_p) + a_i^{(1)} + p_j^{(1)} + d_{i,j}^{(1)} + a_{h(i)}^{(2)} + p_{h(j)}^{(2)} + d_{h(i),h(j)}^{(2)},
\end{equation}
and
\begin{align}
\label{eq:w_ijm}
 w_{i,j,m} = y_{i,j,m}^* - \eta_{i,j} &= \mbox{I}(m=i) (r_{ai}^{(1)} + r_{a,h(i)}^{(2)}) + \mbox{I}(m=j) (r_{pj}^{(1)} + r_{p,h(j)}^{(2)}) \\ \nonumber
& + r_{d |i,j| m}^{(1)} + r_{d |h(i),h(j)| h(m)}^{(2)} + \epsilon_{i,j,m}.
\end{align}
The intercepts $\beta_a$ and $\beta_p$ for actor and partner reports are averaged in (\ref{eq:eta_ij}) to obtain an overall intercept, using the law of total expectation \citep[e.g.][Chapter 4]{casella.berger2002}.  When the sizes of the reporter role groups are unequal, a weighted average can be used.

One possible measure of ``true'' between-individual tie strength is the probability obtained from the structural part of the model $\eta_{i,j}$, setting all reporter effects in the measurement model (\ref{eq:model_measurement}) to zero, leading to $w_{i,j,m}=0$. For a probit model this is defined as  
\begin{equation}
\label{eq:pi_kl_w0}
\pi_{i,j} | (w_{i,j,m}=0) = \Phi (\eta_{i,j}),
\end{equation}
where $\Phi(\cdot)$ is the standard normal CDF.  However, it is well known that for random effects models for binary responses, and other generalised linear mixed models, the magnitude of $\pi_{i,j} | (w_{i,j,m}=0)$ depends on the variance of the random effects that are fixed to zero and such predictions cannot be interpreted as mean predictions (see, for example, \cite{bland.cook2019} for two-level probit and logit models). In the case of a two-level random intercept model, the prediction at a zero value of the random effect can be interpreted as the median probability, but this interpretation does not extend to more complex models with multiple random effects. The difference between (\ref{eq:pi_kl_w0}) and mean predictions may be substantial when the variance of $w_{i,j,m}$ is large.

Mean (marginal) probabilities of individual ties that adjust for reporter effects are obtained by conditioning on $\eta_{i,j}$ and averaging over the random effects in $w_{i,j,m}$.  Letting $\sigma_w^2=\mbox{var}(w_{i,j,m})$ and extending derivations in \cite{bland.cook2019}, it can be shown that the mean probability is given by
\begin{equation}
\pi_{i,j} = \Phi \left ( \frac{\eta_{i,j}}{\sigma_w}  \right )
\end{equation}
(see Section S5 of the supplementary materials for details). An advantage of $\pi_{i,j}$ over $\pi_{i,j} | (w_{i,j,m}=0)$, and other measures of tie strength such as standardised $z_{i,j}$, is that the average probability of a tie from household $k$ to household $l$ based on $\pi_{i,j}$ ($\theta^*_{k,l}$ defined below) is of a comparable magnitude to the sample proportion $p_{k,l}$ in (\ref{eq:p_kl}). The mean probabilities $\pi_{i,j}$ can be viewed as scaled versions of $\pi_{i,j} | (w_{i,j,m}=0)$ and $z_{i,j}$.

From (\ref{eq:model_measurement}), $\sigma_w^2$ depends on $m$ because the actor and partner reporter effects have different variances.  We therefore use the pooled individual reporter variance $\sigma_{r(1)}^2 = 0.5(\sigma_{ra(1)}^2 + \sigma_{rp(1)}^2)$ and household reporter variance $\sigma_{r(2)}^2 = 0.5(\sigma_{ra(2)}^2 + \sigma_{rp(2)}^2)$ which, from the distributional assumptions in (\ref{eq:measurement_dist}) and (\ref{eq:structural_dist}), gives
\begin{equation}
\label{eq:sigma_w}
\sigma_w^2 = \sigma_{r(1)}^2 + \sigma_{r(2)}^2 + \sigma_{rd(1)}^2 + \sigma_{rd(2)}^2 + 1.
\end{equation}
The use of pooled variances applies the law of total variance \citep[e.g.][Chapter 4]{casella.berger2002} to obtain the overall and household reporter variances from the corresponding role-specific variances; as for the intercepts in (\ref{eq:eta_ij}), a weighted average can be used when the numbers of actor and partner reports are unequal.

We use $\pi_{i,j}$ to define model-based density and aggregation measures of the strength of between-household ties, analogs of $p_{k,l}$ and $y_{k,l}^{\mbox{ag}}$ respectively. 

\noindent \textbf{Density approach.} The density approach simply involves taking the average probability of a tie from members of household $k$ to members of household $l$ as
\begin{equation*}
    \label{eq:model_density}
    \theta^*_{k,l} = \frac{1}{n_k n_l} \sum_{i \in S_k, j \in S_l}  \pi_{i,j}.
\end{equation*}
This represents the probability of uniformly randomly selecting an individual $i$ from $S_k$ and an individual $j$ from $S_l$ such that $i$ and $j$ are connected.

\noindent \textbf{Aggregation approach.} Define the vector of predicted probabilities corresponding to the connections between $i \in S_k$ and $j \in S_l$ as $\boldsymbol{\pi}_{k,l}^{(2)} = \left \{ \pi_{i,j}: i \in S_k, j \in S_l \right \}$. The model-based aggregation approach is based on summaries of the random variable $q_{k,l}$, the number of ties from household $k$ to household $l$, where
\begin{align} 
    \label{eq:model_aggregation}
    &q_{k,l} \sim \mathrm{Poisson \ Binomial}\left ( \boldsymbol{\pi}_{k,l}^{(2)} \right ).
\end{align}
The relationship between $\theta^*_{k,l}$ and $q_{k,l}$ can be seen in expectation,
\begin{equation*}
    \theta^*_{k,l} = \frac{\mathbb{E}\left ( q_{k,l} \mid \boldsymbol{\pi}_{k,l}^{(2)} \right )}{n_k n_l}.
\end{equation*}
We can summarise the distribution of $q_{k,l}$ using its CDF,
\begin{equation*}
    \theta^{**}_{k,l}(q) = \mathbb{P}( q_{k,l} \geq q),
\end{equation*}
where $q=1$ gives a model-based representation of the binary \textit{any} edge network $\{y_{k,l}^u\}$ (see Section \ref{subsec:deriving_cluster_network_aggregation}). More generally, we can consider $\theta^{**}_{k,l}(q)$ for a range of $q$ values. For example, to induce greater variation in the strength of connections between households, we can choose the minimum value of $q$ subject to the constraint that
\begin{equation*}
    \mathbb{P} \left ( q_{k,l} \geq q \right ) > 0 \ \mathrm{for \ all} \ k,l \in \left \{1,2,\ldots, H \right \}.
\end{equation*}
In the application that follows, for example, the minimum household size (number of adults) is $2$, and setting $q>4$ results in $\theta^{**}_{k,l}(q)=0$ for a household dyad $(k,l)$ with $n_k=n_l=2$. 

\noindent \textbf{MCMC Estimation.} 
Suppose inference for the model parameters and random effects is based on chains of length $T$, drawn from their posterior distributions.  Denote by $\eta_{i,j}^{(t)}$ the prediction of $\eta_{i,j}$ from substituting the values of the chains for the intercepts and household/individual random effects at MCMC iteration $t$ in (\ref{eq:eta_ij}). Similarly, $\sigma_w^{(t)}$ is the square root of (\ref{eq:sigma_w}) with the variances of the reporter effects replaced by their draws at iteration $t$. Define
\begin{equation}
\label{eq:pi_ij^2}
{\pi}^{(t)}_{i,j} =  \Phi\left( \frac{\eta_{i,j}^{(t)}}{\sigma_w^{(t)}} \right ),
\end{equation}
and
\begin{align*}
\boldsymbol{\pi}_{k,l}^{(2,t)} = \left \{ \pi_{i,j}^{(t)}: i \in S_k, j \in S_l \right \},
\end{align*}
which are then used to derive estimates of $\theta^*_{k,l}$ and $\theta^{**}_{k,l}(q)$:
\begin{align*}
    &\hat{\theta}^*_{k,l} = \frac{1}{T} \sum_{t=1}^T  \sum_{i \in S_k, j \in S_l}   \pi^{(t)}_{i,j}, \\
    &\hat{\theta}^{**}_{k,l}(q) = \frac{1}{T} \sum_{t=1}^T  \mathbb{P}\left (q^{(t)}_{k,l} \geq q \right ),
\end{align*}
where
\begin{equation*}
    q^{(t)}_{k,l} \sim \mathrm{Poisson \ Binomial}\left ( \boldsymbol{\pi}_{k,l}^{(2,t)} \right ).
\end{equation*}

There are several advantages to using these model-based approaches.
First, the model-based predictions $\hat{\theta}_{k,l}^*$ and $\hat{\theta}_{k,l}^{**}$ measure the strength of between-household ties, and can be treated as edge weights in calculation of network statistics such as measures of centrality. 
Second, as the model-based predictions are calculated from the empirical Bayes (or shrinkage) estimates of the household actor, partner and dyad random effects, the predictions are precision-weighted (see Section S7 of the supplementary materials for a discussion of the impact of shrinkage). Third, when the multilevel network model is estimated using MCMC methods, it is straightforward to compute Bayesian credible intervals for $\theta_{k,l}^*$ and $\theta_{k,l}^{**}$, and for network statistics based on them, to indicate uncertainty in their prediction.  Finally, apart from using the multilevel model to derive the household network, estimates of the model parameters, and functions of the parameters, are of substantive interest in their own right; quantities of particular interest include dyadic and generalised reciprocity at the individual and household levels, and variance partitioning coefficients for decomposing the total variance into its constituent parts (including reporter effects).

\section{Application to social support networks in Nicaragua}
\label{sec:application}

\subsection{Data}
\label{subsec:application_data}

Collected in April 2013, the data for this application stem from a survey of adults in a rural community of Indigenous Mayangna and Miskito horticulturalists in Nicaragua. Residents of this community maintain wooden homes, typically built on elevated platforms, with a kitchen area, a porch, and partitioned bedrooms inside the domicile \citep{cimadomo2020documentation}. Households typically centre on monogamously married couples who share the residence with their offspring and other extended kin. Upon forming new relationships, young couples normatively continue to reside with the woman's parents until they have at least a couple of children of their own, at which point they will often elect to build a new home \citep{koster2011hypothetical}. At the time of data collection, the total fertility rate was high, averaging approximately eight births per woman \citep{winking2015fitness}. Meanwhile, elderly individuals often reside in the homes of adult offspring. In terms of livelihood strategies, households are largely self-sufficient, relying on crops from their swidden plots, complemented by foraged foods and domestic animals \citep{koster2013effects}. 

The network survey was administered by one of the co-authors (JK) with the assistance of a translator who was fluent in both Mayangna and Miskito. An earlier census of the community permitted the use of the roster method of eliciting network connections \citep{butts2008social}. The roster included all residents of the community who were at least 18 years old ($n$=108), all but two of whom were interviewed as part of this study. The question that was posed to respondents was designed to encompass the most common types of support that are exchanged among individuals in separate households. Prior to commencing data collection, the question was piloted with approximately 10 participants to finalise the phrasing in the indigenous languages. Specifically, participants were asked, ``Who provides tangible support to you at least once per month?" Relevant domains of support were then listed as examples and included sharing gifts of food or firewood, lending of valuable items such as dugout canoes and axes, and help such as uncompensated assistance clearing agricultural fields or constructing houses. Following the prompt, the translator read aloud the names of the other 107 adult residents (alters) in random order to the participants, who were instructed to verbally respond either affirmatively or negatively depending on the extent to which the named alter met the criteria of the question. Immediately afterward, the inverse question (with the same examples) was posed to the respondents, ``To whom do you provide tangible support at least once per month?" The names of alters were newly randomised and read aloud to participants who registered their responses to each. These questions comprised the first part of an interview that included other components unrelated to social networks. For the overall interview, which lasted approximately one hour, respondents received a modest financial incentive that was roughly equivalent to the standard local wage for five hours of agricultural labour. All interviews occurred privately in a central location, the community's church.

The data collection method presupposes the nomination of individual alters, but the question evokes domains of tangible support that may impact all household members. As examples, gifted foods are cooked and distributed among members of the recipient household, and often multiple household members jointly use a borrowed canoe. Once a material resource has been relinquished, it may be further shared or subdivided in ways that deviate from the donor's intentions or preferences, which has been dubbed the ``targeting problem" in the anthropological literature \citep{hames1987garden}. Furthermore, it is ethnographically common for individuals to prioritise other household members as secondary beneficiaries of support \citep{leonetti2011foundation}. Yet, there can also be opportunities for actors to direct support to specific partners. As an alternative to sending food resources to another household, for example, specific partners could be invited to share a meal in the actor's house. The parallel opportunities for exchanges of social support among individuals and households motivate our multilevel statistical approach.

The 106 interviewees were asked about their exchanges of support with each other and with the two non-respondents who were temporarily absent from the community.  We therefore have data on all 108 $\times$ 107 = 5778 dyads among adult residents.  The full set of four reports are available for all dyads formed among the interviewees, but for dyads involving an interviewee and a non-respondent there are only two reports (on exchanges in each direction from the interviewee's perspective).  After excluding within-household dyads, there are 5594 dyads, 11,188 directed dyads and 21,960 observations. The 208 dyads with only 2 reports are included in the analysis under a missing at random assumption. The 108 adult individuals are nested in 32 households with the number of individuals per household ranging from 2 to 10 (median = 3). The mean age was 34.5 years (SD = 13.8) and there was an equal number of men and women. 

\subsection{Preliminary analysis of dyadic responses}
\label{subsec:application_prelim}

Double-sampling of each tie allows us to assess the extent of agreement between reports from the actor (giver) and partner (receiver) perspectives, and how the correlation between giving and receiving help depends on the role of the reporter.  The sample proportion of positive ties is slightly higher for actor reports (0.239) than for partner reports (0.207) suggesting that, on average, individuals may apply a lower threshold in their reports of giving help than receiving help.  However, the level of agreement at a dyadic level is substantially lower. We examine 
tetrachoric correlations among responses for the same dyad, which are estimates of the correlations among the latent responses $(y_{i,j,m}^*,y_{j,i,m}^*)$ for $m \in \{i,j\}$.  These show only a moderate correlation between actor and partner reports of the same directed tie ($\widehat{\Cor}(y_{i,j,i}^*,y_{i,j,j}^*) = 0.515$). We also find that estimates of the correlation between support given and received (dyadic reciprocity unadjusted for actor and partner effects) differ for the actor and partner perspectives, with $\widehat{\Cor}(y_{i,j,i}^*,y_{j,i,j}^*) = 0.560$ (actor reports) and $\widehat{\Cor}(y_{i,j,j}^*,y_{j,i,i}^*) = 0.429$ (partner reports). This relatively low level of agreement is consistent with findings from other attempts to look at concordance in reporting in other settings \citep[e.g.][]{ready.power2021}. The estimate of reciprocity is considerably higher when based on reports from the same individual ($\widehat{\Cor}(y_{i,j,i}^*,y_{j,i,i}^*) = 0.842$).

\subsection{Research questions}
\label{subsec:application_research_questions}

The multilevel SRM described in Section \ref{sec:multilevel_SRM} is applied to the Nicaraguan data to investigate the following research questions:
\begin{enumerate}[(i)]
\item What proportions of the total variation in reports of between-individual support ties ($y_{i,j,m}^*$) are due to actor, partner and dyadic reporter effects at the individual and household levels?
\item What proportions of the total variation in the true propensity of between-individual support ties ($z_{i,j}$) are due to actor, partner and dyadic effects at the individual and household levels? How do these estimates compare to those from separate analyses of reported support ties from actor and partner perspectives?
\end{enumerate}

We then apply the methods of Section \ref{sec:deriving_cluster_network} to derive household networks from the multilevel SRM to investigate:
\begin{enumerate}[(i)]
\setcounter{enumi}{2}
\item Differences between measures of household centrality from descriptive and model-based aggregation and density approaches to network construction.
\item Differences in measures of household centrality among different network construction methods when dual vs single reports are analysed.
\end{enumerate}

\subsection{Results from multilevel social relations models}
\label{subsec:application_model_results}

The multilevel SRM for double-sampled ties given by (\ref{eq:model_measurement}) and (\ref{eq:model_structural}) was fitted to the Nicaraguan data. The results presented below are the posterior means and 95\% credible intervals for four chains of 3500 MCMC iterations, each using a different set of starting values, with a warm-up sample of 5000 and retaining every second MCMC sample. Convergence was assessed using a range of diagnostics, including visual inspection of trace plots for the multiple chains, tail and bulk effective sample size (ESS), $\hat{R}$ and the number of divergent transitions \citep{vehtari.etal2021}. All diagnostics suggested good convergence with minimum values of 1796 for tail-ESS and 842 for bulk-ESS across all parameters and a maximum $\hat{R}$ of 1.008. Furthermore, trace plots showed adequate mixing for all parameters and increasing the number of MCMC iterations led to little change in the running means of the posterior estimates.  There were no divergent transitions after increasing the target acceptance probability used during the warmup phase from the \texttt{rstan} default of 0.8 to 0.95. 

The fit of the final model was assessed using posterior predictive checks based on replicates of the binary reports of ties $y_{i,j,m}$ simulated from the MCMC parameter chains for the fitted model, including predicted values of all random effects. Test statistics were chosen to assess the model's ability to capture several features of the data: (i) within-dyad correlations, and (ii) standard deviations of out-degree and in-degree across individuals and households, separately for actor and partner reports. For the household statistics, the household network was constructed using the union and density methods of aggregation. In each case, the observed value of the test statistic is close to the posterior predictive mean and lies well within the 2.5 and 97.5 percentiles of the posterior predictive distribution, indicating that the model provides a good fit to these aspects of the data (see Table S8 in the supplementary materials).

Table \ref{tab:measurement} shows parameter estimates from the measurement model of (\ref{eq:model_measurement}). Also shown are estimates of the variance partitioning coefficients (VPCs) indicating the proportions of the total variance in $y_{i,j,m}^*$ implied by the measurement and structural models that are due to the actor, partner and dyadic reporter effects at the individual and household levels (research question (i) in Section \ref{subsec:application_research_questions}). An individual's reporter effect captures their tendency to report in a certain way on exchanges with any other individual, while the household's reporter effect captures clustering in individual reports for members of the same household. As the individual and household reporter effects in (\ref{eq:model_measurement}) depend on whether the reporter $m$ is the actor ($r_{ai}^{(1)},r_{a,h(i)}^{(2)}$) or partner ($r_{pj}^{(1)},r_{p,h(j)}^{(2)}$), with variances of $(\sigma^2_{ra(1)},\sigma^2_{ra(2)})$ and $(\sigma^2_{rp(1)},\sigma^2_{rp(2)})$ respectively, the VPCs differ for the two reports and are displayed in separate columns. There are substantial reporter effects at the individual level with the combined actor/partner and dyad components accounting for 28.7--33.9\% of the total variance (for partner and actor reports respectively).  Individual reporter effects are slightly larger for actors than partners, though an individual's effects when reporting as actor and as partner are highly correlated. Approximately 9\% of the total variance can be attributed to a reporter effect at the individual dyad level, which captures an individual's tendency to report exchanges in either direction with a specific alter. The wide credible intervals for the variances and correlation of the household-level actor and partner effects indicate that they are imprecisely estimated (as there are only 32 households in the sample) and, together with the dyad effect, the household reporter effects account for only 2.6--2.8\% of the total variance. (Recall also that, due to the small number of households, the estimate of the household-level correlation is biased downwards, see discussion of the simulation results in Section \ref{subsec:multilevel_SRM_estimation} and Table S6 in the supplementary materials.) The remainder of the variance in $y_{i,j,m}^*$ is attributed to the residual $\epsilon_{i,j,m}$ (7.3--7.9\% of the total) and the components of $z_{i,j}$ (56.2--60.6\%).

\begin{table}[!htbp]
\centering
\caption{Parameter estimates and variance partitioning coefficients for reporter effects in the measurement model of eq. (\ref{eq:model_measurement}) with 95\% credible intervals}
\label{tab:measurement}
\begin{footnotesize}
\begin{tabular}{lrrrrrr}
\hline
& \multicolumn{2}{c}{} & \multicolumn{2}{c}{VPC:Actor (\%)} & \multicolumn{2}{c}{VPC:Partner (\%)} \\
 \cmidrule(lr){4-5} \cmidrule(lr){6-7}
Parameter  & Est. &  95\% CI & Est. &  95\% CI & Est. &  95\% CI \\
\hline
Actor report intercept $\beta_a$               & -2.14 & (-2.81, -1.45) & & & & \\
Partner-actor report difference $\beta_p - \beta_a$     & -0.48 & (-0.74, -0.23) & & & & \\
Individual: Actor report variance $\sigma^2_{ra(1)}$         & 2.47 & (1.68, 3.51) & 19.2 & (14.3, 24.8) & -- & -- \\
Individual: Partner report variance $\sigma^2_{rp(1)}$         & 3.48 & (2.43, 4.91) & -- & -- & 25.1 & (19.5, 31.4) \\
Individual: Actor-partner report correlation $\rho_{rap(1)}$            & 0.84 & (0.76, 0.90) & & & & \\
Individual: Dyad report variance $\sigma^2_{rd(1)}$         & 1.22 & (0.89, 1.64) & 9.5 & (7.6, 11.6) & 8.8 & (7.0, 10.8) \\
Household: Actor report variance $\sigma^2_{ra(2)}$         &  0.09 &  (0.0001, 0.42) & 0.7 & (0.001, 3.2) & -- & -- \\
Household: Partner report variance $\sigma^2_{rp(2)}$         & 0.08 & (0.0001, 0.42) & -- & -- & 0.6 & (0.0004, 3.7) \\
Household: Actor-partner report correlation $\rho_{rap(2)}$            & 0.17 & (-0.74, 0.88) & & & & \\
Household: Dyad report variance $\sigma^2_{rd(2)}$         & 0.27 & (0.15, 0.43) & 2.1 & (1.2, 3.3) & 2.0 & (1.1, 3.0) \\
Residual: $\sigma^2_\epsilon$ & 1$^\dagger$ & -- & 7.9 & (6.4, 9.5) & 7.3 & (5.9, 8.8) \\
\hline
Total VPC & -- & -- & 39.4 & (34.5, 44.7) & 43.8 & (38.6, 49.4) \\
\hline
\multicolumn{7}{p{0.9\textwidth}}{$^\dagger$\footnotesize{The residual variance, $\sigma^2_\epsilon = \mbox{var}(\epsilon_{i,j,m})$ is constrained to 1 for identification.}}
\end{tabular}
\end{footnotesize}
\end{table}

To assess the impact of adjusting for reporter error (research question (ii)) the results from the structural model of (\ref{eq:model_structural}), fitted to the double-sampled data (dual reports), were compared to those from fitting separate models to the actor reports ($y_{i,j,i}$) and partner reports ($y_{i,j,j}$).  These single-report models have the same form as (\ref{eq:model_structural}) with $z_{i,j}$ replaced by the underlying latent responses $y_{i,j,i}^*$ and $y_{i,j,j}^*$ and an identification constraint on the residual variance, $\mbox{var}(d_{i,j}^{(1)})=1$. Table \ref{tab:recip.vpc} shows estimates of the VPCs, defined as the percentage of the total variance in $z_{i,j}$ (or $y_{i,j,m}^*$ in the case of the single-report models) explained by each component of variation, and the individual and household-level generalised and dyadic reciprocity correlations.

\begin{table}[!htbp]
\centering
\caption{Estimates of variance partitioning coefficients and reciprocity correlations from the structural model for dual reports and comparable models for single reports with 95\% credible intervals}
\label{tab:recip.vpc}
\begin{footnotesize}
\begin{tabular}{lrcrcrc}
\hline
& \multicolumn{2}{c}{Dual reports} & \multicolumn{2}{c}{Actor reports} & \multicolumn{2}{c}{Partner reports} \\
\cmidrule(lr){2-3} \cmidrule(lr){4-5} \cmidrule(lr){6-7}
  & Est. &  95\% CI & Est. &  95\% CI & Est. &  95\% CI \\
\hline
\emph{Variance partitioning coefficients (\%)}  & & & & & & \\
Individual: Actor $a_i^{(1)}$               & 17.5 & (12.7, 22.8) & 20.4 & (15.1, 26.7) & 11.5 & (8.1, 15.8) \\
Individual: Partner $p_j^{(1)}$             & 16.2 & (11.9, 21.1) & 11.4 & (8.1, 15.6) & 23.7 & (17.8, 30.5) \\
Individual: Dyad $d_{i,j}^{(1)}$            & 18.8 & (15.3, 22.4) & 30.6 & (26.7, 34.4) & 29.4 & (25.6, 33.2) \\
Household: Actor $a_{h(i)}^{(2)}$           & 2.7  & (0.02, 8.5) & 3.0 & (0.04, 8.8) & 4.2 & (0.6, 9.6) \\
Household: Partner $p_{h(j)}^{(2)}$         & 2.8  & (0.05, 8.3) & 4.0 & (0.5, 9.2) & 2.8 & (0.02, 9.7) \\
Household: Dyad $d_{h(i),h(j)}^{(2)}$    & 41.9 & (35.2, 48.7) & 30.7 & (25.7, 35.8) & 28.4 & (23.7, 33.2) \\
& & & & & & \\
\emph{Reciprocity correlations}  & & & & & & \\
Individual: Generalised $\rho_{ap(1)}$  & 0.95 & (0.91, 0.98) & 0.33 & (0.11, 0.52) & 0.18 & (-0.05, 0.40) \\
Individual: Dyadic $\rho_{dd(1)}$       & 0.90 & (0.80, 0.98) & 0.37 & (0.30, 0.44) & 0.39 & (0.30, 0.47) \\
Household: Generalised $\rho_{ap(2)}$   & 0.45 & (-0.57, 0.92) & 0.44 & (-0.41, 0.91) & 0.19 & (-0.62, 0.83) \\
Household: Dyadic $\rho_{dd(2)}$        & 0.98 & (0.96, 0.99) & 0.93 & (0.89, 0.97) & 0.84 & (0.77, 0.90) \\
\hline
\end{tabular}
\end{footnotesize}
\end{table}

We first note that estimates from the respective single-report models are broadly similar, with the notable exception of the individual-level actor and partner VPCs.  Although the sum of these VPCs is similar (31.8\% and 35.2\% for actor and partner reports respectively), the decomposition of the variance depends on the reporter's role in a dyad: when individuals are reporting as actors, the variance in their reports of help given is almost double the variance in their reports of help received, but this pattern is reversed for partner reports. After adjusting for reporter effects using the dual-report model, however, the variance components for individual actors and partners are approximately equal (explaining a total of 33.7\% of the variance).  In contrast, the proportions of variance attributed to household-level actor and partner effects are similarly small for all models and, as for the variances of the household-level reporter effects in the measurement part of the dual-report model, imprecisely estimated.

Comparing dual-report and single-report estimates of the VPCs for dyad effects, the most striking difference is the partial shift in variance from the individual level to the household level after adjusting for reporter effects.  This pattern is consistent with the strong individual reporter dyad effects from the measurement part of the dual-report model (see Table \ref{tab:measurement}) which in the single-report models is absorbed by the individual dyad effects $d_{i,j}^{(1)}$ leading to an inflated individual-level dyad variance (accounting for approximately 30\% of the variance when actor and partner reports are analysed separately versus 19\% in the dual-report model). After adjusting for reporter effects, the variance attributable to household dyads increases from 28--31\% to 42\%.

Tables \ref{tab:measurement} and \ref{tab:recip.vpc} show the proportions of variance attributed to each random effect in the measurement and structural model respectively.  An alternative way to quantify the relative contributions of the different components of variation is to compute estimates of the correlations between pairs of responses $(y_{i,j,m},y_{i^\prime,j^\prime,m^\prime})$ that share individuals or household members.  Expressions for these correlations, based on parameters of the model of (\ref{eq:model_measurement}) and (\ref{eq:model_structural}), and estimates for the Nicaraguan data, are given in the supplementary material (Section S2).  For example, the correlation between an individual $i$'s reports of giving help to two different individuals $j$ and $j^\prime$, $\mbox{cor}(y_{i,j,i},y_{i,j^\prime,i})$, is estimated as 0.322, which reflects a combination of actor and reporter effects. This correlation increases to 0.614 when the partners/recipients $j$ and $j^\prime$ are in the same household, due mainly to the large contribution from the household dyad variance, $\mbox{var}(d_{h(i),h(j)}^{(2)})$, noted above. When reporter effects are absent (i.e. when reports are from two different individuals), there is a marked decrease in these correlations to 0.114 and 0.391 respectively.  

Turning to reciprocity, estimates of individual-level generalised and dyadic reciprocity are substantially smaller for the single-report models, a pattern consistent with attenuation bias due to measurement (reporter) error.  After adjusting for reporter effects, both correlations are close to 1.  A generalised reciprocity of 1 implies that the SRM could be simplified by replacing the individual actor and partner effects by a common individual effect, representing an individual's tendency to participate in exchanges in either direction.  A dyadic reciprocity close to 1 implies that help given by any individual $i$ to any individual $j$ is almost always reciprocated. In other words, there were very few bilateral reports of unidirectional assistance within the dyad beyond those reflected in other effects in the model. Meanwhile, at the household level, both the dual and single report models suggest little evidence of generalised reciprocity as the 95\% credible intervals are very wide and include zero.  However, it is important to note that our simulations showed that, as for the correlation between actor and partner reporter effects at the household level, this parameter is biased downwards and its SE is biased upwards for our data structure (see Table S7 in the supplementary materials). Finally, estimates of household-level dyadic reciprocity are high for all models.

Substantively, the results of the structural model are seemingly the result of a kin-based social organisation and the framing of the social support questions that were posed to respondents. First, most adult residents have close kin (parents, offspring, or siblings) residing in other households, and these kinship ties help to form the basis of social organisation. When individuals cooperate and associate with residents of other households, it is very common for these partners to be from households with a close kinship tie to the actor's household \citep{koster2018}. These kin-based interactions likewise extend to affinal kin (i.e. in-laws), which helps to explain why the household-level variances and covariances are so pronounced. That is, when there is a close kinship tie between two respective members of households $k$ and $l$, then it is likely that bidirectional support is exchanged among all of the adults in the households.

Second, the high reciprocity in the responses partly stems from the framing of the questions, both in terms of the open-ended time frame (i.e. who helps at least once per month?) and the multiple domains of support, which encompass a number of different ways that households exchange support. This framing presumably elicits general propensities to provide assistance rather than recall of specific helping behaviours. An alternative approach is to ask about detailed types of support over briefer intervals, such as asking about specific borrowed items or food sharing in the past week \citep[e.g.][]{kasper2015helps}. Almost inevitably, such framings yield responses that exhibit less dyadic reciprocity, which in turn may not be representative of longstanding dyadic relationships. Ethnographically, flows of assistance in the present study community appear to be bidirectional over long intervals, though the binary response employed in this study potentially obscures imbalances in the flow of support within dyads. It is also worth noting that the reciprocity correlations are estimated from random effects that can change magnitude and sign when covariates are included, and it is likely that the strength of the correlations would be attenuated by the inclusion of variables for dyadic kinship and individuals' demographic attributes.

\subsection{Between-household networks}
\label{subsec:application_network_analysis}

In Section \ref{sec:deriving_cluster_network}, we considered the aggregation and density approaches to constructing a household network from individual-level data.  The simple forms of these approaches based on descriptive statistics on the household dyad are, respectively, the union method (a binary indicator of the presence of any tie from household $k$ to household $l$, $y_{k,l}^u$) and the sample proportion of ties from $k$ to $l$ ($p_{k,l}$).  Their model-based generalisations, derived from the fitted multilevel SRM, are $\hat{\theta}^{**}_{k,l}(q)$ and $\hat{\theta}^*_{k,l}$ respectively (defined in Section \ref{subsec:deriving_cluster_network_SRM}).  In this section, we study the networks produced by these methods when applied to the individual-level social support network analysed above to investigate research questions (iii) and (iv).  We begin with an analysis of the binary union indicator used in most previous research, comparing the network from pooling the double-sampled data with those based on either actor or partner reports. We next compare the descriptive and model-based versions of the aggregation and density networks, and finally compare the two model-based networks. (As we have an almost complete census of adults in the 32 households, the caveats about household sampling schemes noted in Section \ref{subsec:deriving_cluster_network_aggregation} for the union and model-based aggregation methods do not apply here.) 

Differences between the methods can be elucidated by exploring node (household)-level measures of centrality, which capture the importance of a household to network structure \citep{butts2008social}. Generally, these measures capture local or global notions of importance. Local centrality measures include in- and out-degree, which are sums of the inward and outward connections (respectively) associated with a node. Instead of treating connections equally when summing, eigenvector centrality incorporates global structure by attaching greater weight to connections to more central nodes \citep{bonacich2007some}. Lastly, betweenness centrality is a global measure that quantifies the proportion of shortest paths that go through a node. Further details of the calculation and interpretation of network measures can be found in \cite{saxena2020centrality}. 
Centrality measures are useful for quantifying the ``social capital" of households in a social network. Because social connections can disseminate economic opportunities and resources \citep{jackson2022inequality, redhead2024social}, household-level differences in centrality can be indications of differences in the possibility of economic stability and mobility. Additionally, centrality has been suggested as a tool for seeding interventions; e.g., starting dynamic processes (or stemming them) at central nodes may facilitate a faster (slower) spread across network connections \citep{banerjee2014gossip}. Therefore, it is informative to compare the centrality measures implied by the aggregation and density methods of household network construction.

\subsubsection*{Binary ``any tie'' networks (the union method)}

When data are double-sampled, binary indicators of household ties can be obtained by pooling both reports ($y_{k,l}^u$) or by selecting either actor or partner reports of each directed tie (denoted by $y_{k,l}^{u(a)}$ and $y_{k,l}^{u(p)}$ respectively).  As noted in Section \ref{subsec:application_prelim}, there is a low level of agreement between actor and partner reports of the same tie, which leads to a higher edge prevalence for the pooled network ($83\%$) compared to the single-report networks ($70\%$ and $67\%$ for actor and partner reports respectively). Thus, when reports are combined, the network is almost fully connected and the between-household variation in centrality measures is lower than when one report is used (for example, the coefficient of variation (CV) for betweenness is $54\%$ for combined reports versus approximately $80\%$ for single reports). In spite of the inconsistency in actor and partner reports, the correlation between household centrality estimates for the two networks is high for betweenness (0.80) and eigenvector centrality (0.91).  For degree centrality, however, the correlation is much lower (0.32 for out-degree and 0.43 for in-degree). Moreover, for out-degree the CV is much higher for actor than for partner reports ($30\%$ vs $21\%$), while the pattern is reversed for in-degree ($24\%$ vs $37\%$).  Taken together, these results suggest that reporter effects inflate the between-household variance in the number of outward ties based on actor reports and in the number of inward ties based on partner reports, which is consistent with the patterns in the proportions of variance due to individual actor and partner effects estimated from the multilevel SRMs fitted separately to actor and partner reports (see Table \ref{tab:recip.vpc}).

This analysis illustrates some of the drawbacks with simple aggregation methods.  On the one hand, combining reports inflates edge prevalence which masks differences between households in measures of their centrality while, on the other hand, networks based on either an actor or partner perspective are especially subject to distorting reporter effects. Reducing individual-level data to `any tie' binary indicators is also wasteful as the number of individual ties between households (or proportion) captures information on tie strength. 

\subsubsection*{Comparison of descriptive and model-based networks}

We next compare centrality estimates for household networks obtained from simple descriptive methods (using $y_{k,l}^u$ and $p_{k,l}$) and their model-based generalisations ($\hat{\theta}^{**}_{k,l}(q)$ and $\hat{\theta}^*_{k,l}$ respectively). For the comparison of the aggregation approaches we set $q=1$ to obtain $\hat{\theta}^{**}_{k,l}(1)$, the estimated probability that at least one adult in household $k$ gives support to at least one adult in household $l$, which is the model-based counterpart of the binary `any tie' indicator $y_{k,l}^u$.

For all centrality measures, there are moderate to high correlations between estimates from the descriptive and corresponding model-based aggregation networks (0.73 for betweenness, 0.82 for eigenvector, 0.87 for out-degree and 0.82 for in-degree).  Although there is broad agreement in the centrality estimates on each measure, especially at the extremes, there are some notable discrepancies (see scatterplots in Figure S1).  For betweenness, with the lowest correlation, model-based aggregation results in substantially greater between-household variation than the binary `any tie' approach (CV = 132\% vs 54\%) which suggests that the use of valued edges leads to better discrimination between households.  Centrality estimates based on the descriptive and model-based density networks (using $p_{k,l}$ and $\hat{\theta}^*_{k,l}$) are highly correlated for each measure, including betweenness (above 0.8; see also Figure S2), and have similar between-household variances.  This suggests that betweenness is especially sensitive to whether binary or valued edges are used. In further analysis, we find that differences between $p_{k,l}$ and $\hat{\theta}^*_{k,l}$ depend on the number of observations in the household dyad $(k,l)$, which is consistent with shrinkage in the household dyad random effect that dominates $\hat{\theta}^*_{k,l}$.  However, this does not imply that household centrality estimates based on $\hat{\theta}^*_{k,l}$ will be subject to shrinkage (see Section S7.3 for further discussion).

If we consider comparisons between networks based on $\hat{\theta}^{**}_{k,l}(1)$ and $y_{k,l}^{u(a)}$ or $y_{k,l}^{u(p)}$ (rather than the pooled $y_{k,l}^u$), an interesting pattern emerges for the degree measures. While degree for the model-based network broadly aligns with degree for the pooled `any tie' network, the extent of agreement when using either actor or partner reports differs for in-degree and out-degree.  For in-degree, there is close agreement between the model-based network and that using $y_{k,l}^{u(a)}$ (correlation of 0.96), but large discrepancies emerge when the union network is based only on partner reports $y_{k,l}^{u(p)}$ (correlation of 0.54). The reverse is observed for out-degree with correlations of 0.59 and 0.89 between the $\hat{\theta}^{**}_{k,l}(1)$ network and the actor and partner union networks respectively (see also scatterplots in Figure S3). The same patterns are observed in the comparison of degree measures for the density networks based on the model relative to  sample proportions from actor and partner reports (see Figure S4). These findings are in line with those noted above for the comparison of degree measures based on the union method for pooled versus single reports, and demonstrate the advantages of double-sampled data analysed using principled statistical methods.

\subsubsection*{Comparison of model-based networks}

Our final comparison is of the two model-based networks. For the aggregation approach, we now set $q=4$ to obtain $\hat{\theta}^{**}_{k,l}(4)$ which is the estimated probability of at least four ties from household $k$ to household $l$ (the minimum number of possible individual ties between households as all have at least two respondents). Changing from $q=1$ to $q=4$ decreases edge prevalence from $78.7\%$ to $28.2\%$ (compared with $27.3\%$ for the density approach), and increases variation between households in their strength of connections.

\begin{figure} [!htbp]
    \centering
    \includegraphics[width=0.8\linewidth]{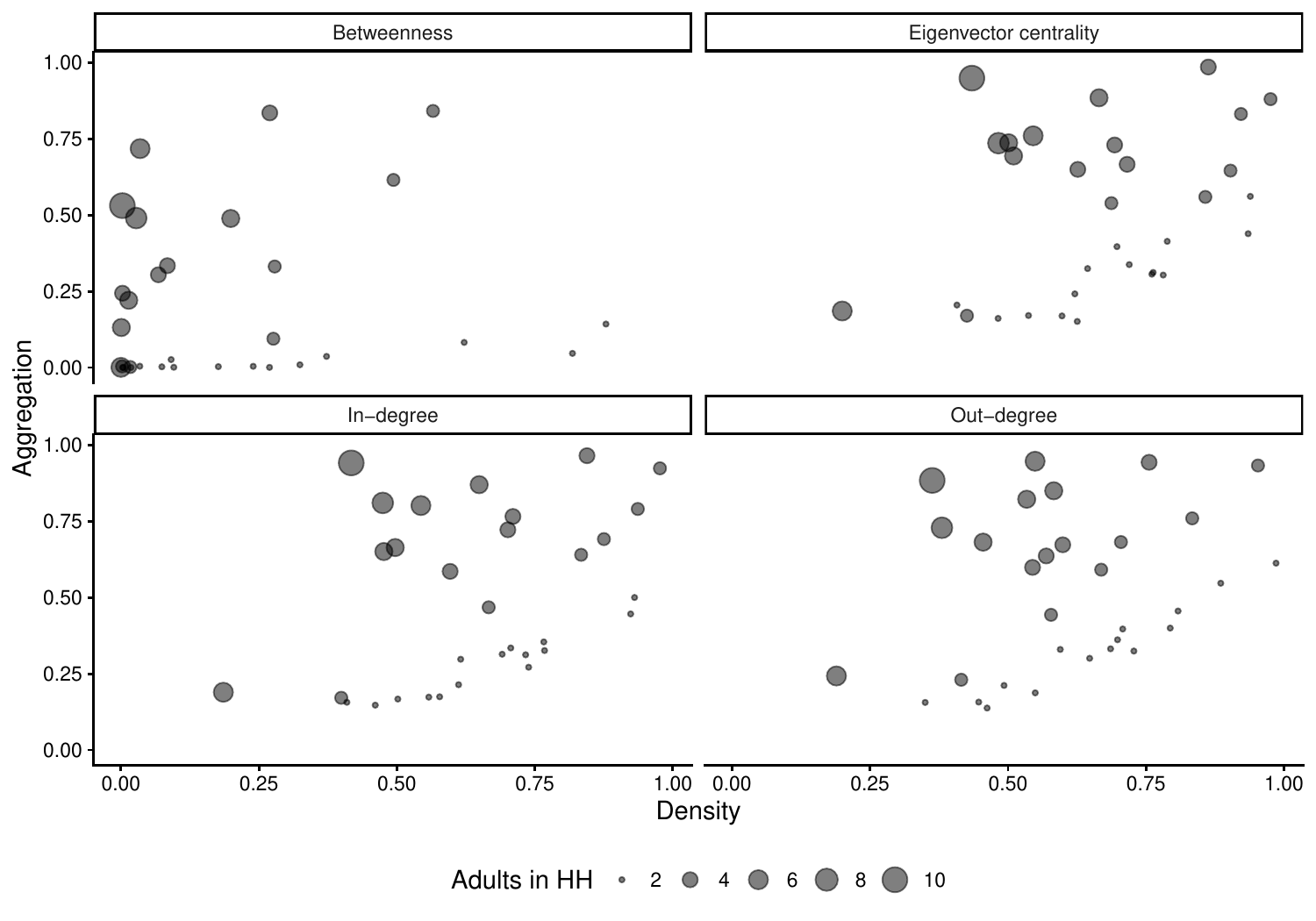}
    \caption{Normalised centrality measures, in which each measure is divided by its maximum value, for the household (HH) networks specified by the model-based aggregation ($q=4$) and density methods.}
    \label{fig:hh_centrality}
\end{figure}

We expect that the household network from the aggregation method will exhibit centrality measures that prioritise larger households (where household size is the number of adults in the household) more than the density method.  In Figure~\ref{fig:hh_centrality}, we see that the difference between the eigenvector centrality of the network constructions aligns with this intuition. Indeed, the difference between the two networks is largely driven by adult household size. Among households with a similar number of adults, the aggregation and density estimates of eigenvector centrality are highly correlated (increasing from 0.32 for all 32 households, to 0.70 for the subset of 25 households with 2--4 adults, and 0.85 for the subset of 15 households with 2 adults). Additionally, the weighted in- and out-degrees resemble eigenvector centrality (see Figure \ref{fig:hh_centrality}), and so follow the same trend. 

A version of Figure \ref{fig:hh_centrality} with the addition of 95\% credible intervals is given in Figure S5 of the supplementary materials. For each MCMC draw $t$, the household-level edge weights for the aggregation and density methods ($\theta_{k,l}^{**(t)}(4)$ and $\theta_{k,l}^{*(t)}$) are computed from the individual-level edge weights $\pi_{i,j}^{(t)}$ given by (\ref{eq:pi_ij^2}). Posterior samples of the four centrality measures are then obtained from the draws of the household edge weights.  For both methods, the credible intervals are markedly wider for betweenness than for the other centrality measures.  For betweenness, edge weights are converted to binary indicators of the shortest path between pairs of households, so small changes in edge weights across MCMC draws can lead to a large amount of variation in the shortest paths.  The other centrality measures are linear functions of the edge weights and are therefore more stable across MCMC draws.

\section{Discussion}
\label{sec:discussion}

We have proposed an extension of the widely-used Social Relations Model for the analysis of clustered social network data containing individuals' reports on their ties with individuals in different clusters, where data are ``double-sampled'' such that each member of an individual dyad reports on ties in both directions. While previous research on clustered networks has focused on settings where ties can occur only within clusters, we consider the case of between-cluster ties where the between-cluster network is of central interest.  Our model-based approach offers several advantages over simple aggregation methods that combine dual reports of the same tie across individuals and clusters to obtain binary indicators of between-cluster ties. The proposed multilevel SRM quantifies the extent of reporting errors at the individual and cluster levels, and yields estimates of actor, partner and dyad effects at each level that are adjusted for reporter effects. Measures of between-cluster ties can then be derived from the fitted model. We propose two such measures, both adjusted for reporting errors and using the individual-level data to estimate the strength, rather than simply the binary presence, of ties: the predicted probability of more than $q$ individual ties between clusters (referred to as ``aggregation'') and the predicted probability of a between-cluster tie (``density''). We discuss the choice between these measures and demonstrate how they can be used as inputs in analysis of the cluster-level network, for example to obtain estimates of centrality measures for each cluster in the network.

For social and behavioural scientists, several insights about research design are evident. First, among scholars of social network analysis, it has long been known that informants' reports almost invariably entail inaccuracies, which can bias inferences about statistical parameters \citep{feld2002detecting}. As seen in Table \ref{tab:recip.vpc} , the present analysis contributes to that literature by showing that the structural component of the dual report model and the single report models differ substantially in the partitioning of variation and reciprocity correlations at both the individual and household levels. Because of the opportunity to partition reporter effects, which is not possible with single reports, there are clear merits to collecting multiple data points on each directed dyad, as with double-sampled data. As a modest caveat to that generalisation, as the proportion of variance due to the individual- and household-level dyad reporter effects increases, an implication is that there are fewer opportunities for respondents to agree on unidirectional assistance between individuals and households, with concomitant implications for estimates of reciprocities in the structural model. It is typical for data collectors to pose the double-sampled questions in immediate succession, as in the present data from Nicaragua, but it would be helpful for researchers to examine the sensitivity of these dyad report variances to the lengths of the temporal intervals between reporters' responses to the respective questions.

Other important aspects of study design are the number of households and number of observations per household.  In the Nicaraguan study, the number of households is relatively small which leads to large credible intervals for the variances (and associated variance partitioning coefficients) and correlations of the actor and partner household effects, and downward bias in the correlations. We recommend that researchers carry out their own simulations to assess whether all model parameters are estimable for their data structure.  In the case of a small number of households, the household-level component of the measurement and structural models could be simplified by reducing the bivariate role-specific effects to univariate effects.  In general, the width of credible intervals for predictions of the household-level actor and partner effects depends on the number of observations per household, with more information on a household leading to more precise predictions. This in turn affects credible intervals for predicted probabilities of between-household ties and summaries of the household network (such as household centrality measures). While double-sampling effectively doubles the number of observations on each household, more extensive reports on the existence of a relationship can be obtained using ``cognitive social structure'' approaches \citep{krackhardt.1987,feltham.etal2025} where respondents report on many dyadic relationships beyond just those they are involved in. For example, in a recent study \citep{feltham.etal2025} respondents were asked to assess the existence of various types of ties for a sample of up to 40 individual dyads (not involving the respondent), with each dyad assessed by 4.43 respondents on average. Another way to increase the amount of information per household, and thus the precision in estimates of the household network, is to consider ``multiplex'' or ``multilayer'' networks \citep{kivela.etal14} where individuals report on exchanges of multiple kinds of support (e.g. financial, practical and advice).

The proposed multilevel SRM can be extended to accommodate multiple reports of ties and multilayer networks. While we consider dual reports of directed ties $i \rightarrow j$ where reporter $m \in \{i,j\}$, it is straightforward to adapt the measurement model to the more general case of multiple reports where reporter $m \notin \{i,j\}$. Such designs can be exploited to study reporter effects, although actor and partner reporter effects can only be distinguished when $m \in \{i,j\}$. Multilayer networks lead to multivariate dyadic data. In such cases, the dIRT model of \cite{gin.etal2020} can be extended to jointly model the layers using a multilevel multivariate SRM, allowing for correlations between layer-specific actor, partner and dyad effects at the individual and cluster levels; where between-layer correlations are high, `overall' or summary random effects may be specified with different weights or loadings for each layer in an SRM extension of a multilevel factor model. The measurement model can also be generalised to handle other types of response, such as counts and continuous measures of tie strength, and mixtures of different response types in the case of multilayer networks.  A major advantage of MCMC estimation is its flexibility in allowing these and other extensions to the model specification presented here.

Another generalisation is to include covariates in the measurement or structural models to determine the extent to which the components of variation can be explained by observed attributes.  Potential covariates in the measurement model, which could explain unexplained reporter tendencies, include characteristics of the reporter (e.g. gender) and interactions between their characteristics and those of the other individual and their household (e.g. to assess whether any effect of the reporter's gender depends on the gender of the alter or the distance to the alter's household). Covariates in the structural model may be characteristics of the actor and partner, their respective households, and the individual and household dyad.  For example, Koster's \citeyear{koster2018} analysis of the same data (with separate models for actor and partner reports) considers covariates relating to individuals (e.g. gender and age), households (wealth), individual dyads (measures of relatedness) and household dyads (e.g. distance and average relatedness). In contrast to single report models, an advantage of double-sampled data is that the effects of covariates can be disentangled from reporter effects.

We show that the possibilities for reporter effects to bias inferences are compounded when using simple descriptive methods, such as the union method, to construct household-to-household networks. For researchers who wish to use estimates of centrality measures in subsequent analyses, the discrepancies that are evident in the supplemental figures imply potentially large differences as a result of choices about network construction. The discrepancies are particularly evident in comparisons of networks based on double-sampled vs single reports, but Section S7.3 and Figure S7 illustrate a household that changes from the top quartile to the bottom quartile of eigenvector centrality when changing from the union of double-sampled reports to the corresponding model-based aggregation approach. Such comparisons illustrate the sensitivity of centrality measures to assumptions about network construction. Approaches that tacitly assume the absence of reporter effects are difficult to justify amid the broader literature on the ubiquity of these effects \citep{feld2002detecting}, while reducing individual-level data to binary indicators of household ties is wasteful.

Whereas the merits of the model-based approaches to simple descriptive methods are apparent, practitioners may desire guidance on the choice between the model-based aggregation and model-based density methods. Provided that the sampling requirements for both methods are met, however, it remains open-ended which of the alternatives is more appropriate for a particular application, and practitioners should be empowered to select between the aggregation and density methods based on the characteristics of the measured network variable and the analytical goals. Ethnographically, for instance, the density method seems moderately more appropriate for the household-to-household network in the Nicaraguan study community. In a qualitative comparison of the estimated probabilities of household ties ($\hat{\theta}^{**}_{k,l}(4)$ for the aggregation method and $\hat{\theta}^{*}_{k,l}$ for the density method), the density method better accords with patterns of social support that are observable with other methods, such as a yearlong study of food sharing in the same community \citep{kosterleckie2014}. Although the aggregation method generally yields comparable estimates for many of the household dyads, there are also cases that are counterintuitive ethnographically. In most of these cases, the discrepancies seemingly arise from considerations of household demographics, where discernibly strong inter-household ties are evidently underestimated by the aggregation method because one or both of the households had a small number of adults. As seen in Figure 1, differences between the density and aggregation methods are driven primarily by adult household size, and researchers can leverage their qualitative understandings of the dyadic relationships to make an appropriate choice.

Finally, we have shown how the model presented here can be used to construct a household-to-household network. However, 
depending on the type and distribution of resources, many of these reported individual-to-individual (I-I) ties would be better characterised as individual-to-household (I-H) or household-to-household (H-H) relationships. 
That is, food is often a household good, making its exchange a H-H relationship. 
Childcare, on the other hand, may be considered an I-H relationship between a caregiver and the household of the children being cared for. In this paper, we assume that individuals are reporting on I-I relationships. Accounting for H-H and I-H relationships would require different measurement models and alternative methods for constructing the household network (discussed in Sections~\ref{subsec:multilevel_SRM_measurement} and \ref{subsec:deriving_cluster_network_aggregation} respectively). Developing a model for simultaneously representing I-I, I-H, and H-H relationships is a goal for future research.

\section*{Funding and acknowledgments}
Funding for the data collection in Nicaragua was provided by the Charles P. Taft Research Center and the Leakey Foundation. 
The authors gratefully acknowledge funding from two UK Research \& Innovation grants: ES/V006495/1 (Developing Latent Hierarchical Network Models for Cross-Cultural Comparisons of Social and Economic Inequality; E.P.) and UKRI3351 (A Cross-Cultural Study of Variability in Economic Prosperity using Machine-Learning and Statistical Analysis; E.P. and F.S.).

\bibliographystyle{apacite}
\bibliography{References}

\section*{Appendix: Key notation}

\begin{table}[!htbp]
\centering
\begin{footnotesize}
\begin{tabular}{ll}
\hline
Term & Description \\
\hline
\multicolumn{2}{l}{\textbf{Data structure}} \\ 
$y_{i,j,m}$ & Observed binary response for whether individual $i$ helps $j$, reported by individual $m \in \{i,j\}$ \\
$y_{i,j,m}^*$ & Latent variable underlying $y_{i,j,m}$ such that $y_{i,j,m}=\mbox{I}(y_{i,j,m}^*>0)$ \\
$h(i)$, $h(j)$, $h(m)$ & Indices for the household of individuals $i$, $j$ and $m$ \\
$k$, $l$ & Alternative indices for households: actor $k$ and partner $l$ \\ [6pt]
\multicolumn{2}{l}{\textbf{Measurement model}} \\ 
$z_{i,j}$ & Latent variable representing true propensity that individual $i$ helps individual $j$ \\
$r_{ai}^{(1)}$, $r_{a,h(i)}^{(2)}$ & Random effects for tendencies of individual $i$ and household $h(i)$ to report giving help \\
$r_{pj}^{(1)}$, $r_{p,h(j)}^{(2)}$ & Random effects for tendencies of individual $j$ and household $h(j)$ to report receiving help \\
$r_{d|i,j|m}^{(1)}$ &  Individual dyad reporter effect (allows correlation in reports by $m$ of exchanges with a specific person)\\
$r_{d|h(i),h(j)|h(m)}^{(2)}$ &  Household dyad reporter effect (allows correlation in reports by $m$'s hh of exchanges with a specific hh)\\
$\epsilon_{i,j,m}$ & Residual specific to $m$'s report of help given by $i$ to $j$ \\ [6pt]
\multicolumn{2}{l}{\textbf{Structural model}} \\ 
$a_i^{(1)}$, $a_{h(i)}^{(2)}$  &  Random effects for tendencies of individual $i$ and household $h(i)$ to give help \\
$p_j^{(1)}$, $p_{h(j)}^{(2)}$   & Random effects for tendencies of individual $i$ and household $h(i)$ to receive help \\
$d_{i,j}^{(1)}$ & Individual dyad effect specific to help given by $i$ to $j$ \\
$d_{h(i),h(j)}^{(2)}$ & Household dyad effect specific to help given by $h(i)$ to $h(j)$ \\ [6pt]
\multicolumn{2}{l}{\textbf{Household network}} \\ 
$p_{k,l}$ & Sample proportion of helping ties from household $k$ to household $l$ \\
$y_{k,l}^{u}$ & An indicator of the presence of a tie from household $k$ to household $l$ \\
$y_{k,l}$ & A count of the ties from household $k$ to household $l$ \\
$\theta_{k,l}^*$ & Average probability of a helping tie from household $k$ to household $l$\\
$q_{k,l}$ & Random variable for number of individual helping ties from household $k$ to household $l$ \\
$\theta_{k,l}^{**}(q)$ & $\mathbb{P}( q_{k,l} \geq q)$, the probability that the total number of ties from household $k$ to $l$ is above $q$ \\
\hline 
\end{tabular}
\end{footnotesize}
\end{table}

\makeatletter
\renewcommand \thesection{S\@arabic\c@section}
\renewcommand\thetable{S\@arabic\c@table}
\renewcommand \thefigure{S\@arabic\c@figure}
\makeatother

\pagebreak
\begin{center}
\LARGE Supplementary Material
\end{center}

\setcounter{section}{0}

\section{Data structure for multilevel Social Relations Model}

Table \ref{tab:ind-level-re-IDs} shows an example of the identifiers for the individual-level random effects in the measurement and structural models.  In this example, we consider the data structure for the individual dyads formed between the members of two households, H1 and H2, each with two individuals (with IDs 1 and 2 for members of H1, and 3 and 4 for members of H2).  Recall that the reporter effects $r_{ai}^{(1)}$ and $r_{pj}^{(1)}$ are defined only for actor and partner reports respectively.

\begin{table}[!htbp]
\centering
\caption{Example of identifiers for individual-level random effects.  There are four individual dyads for between-household ties for individuals 1 and 2 in household H1 and individuals 3 and 4 in household H2, and four records per dyad for actor and partner reports of directed ties.}
\label{tab:ind-level-re-IDs}
\begin{footnotesize}
\begin{tabular}{ccccccccc}
\hline
 &  &  & \multicolumn{6}{c}{Identifiers for random effects} \\
\cline{4-9} 
Actor & Partner & Reporter & & & & & & \\
$i$ & $j$ & $m$ & $r_{ai}^{(1)}$ & $r_{pj}^{(1)}$ & $r_{d|ij|m}^{(1)}$ & $a_i^{(1)}$ & $p_j^{(1)}$ & $d_{i,j}^{(1)}$\\
\hline
1 & 3 & 1 & 1 & $-$ & 1 & 1 & 3 & 1 \\
1 & 3 & 3 & $-$ & 3 & 2 & 1 & 3 & 1 \\
3 & 1 & 1 & $-$ & 1 & 1 & 3 & 1 & 2 \\
3 & 1 & 3 & 3 & $-$ & 2 & 3 & 1 & 2 \\
\hdashline
1 & 4 & 1 & 1 & $-$ & 3 & 1 & 4 & 3 \\
1 & 4 & 4 & $-$ & 4 & 4 & 1 & 4 & 3 \\
4 & 1 & 1 & $-$ & 1 & 3 & 4 & 1 & 4 \\
4 & 1 & 4 & 4 & $-$ & 4 & 4 & 1 & 4 \\
\hdashline
2 & 3 & 2 & 2 & $-$ & 5 & 2 & 3 & 5 \\
2 & 3 & 3 & $-$ & 3 & 6 & 2 & 3 & 5 \\
3 & 2 & 2 & $-$ & 2 & 5 & 3 & 2 & 6 \\
3 & 2 & 3 & 3 & $-$ & 6 & 3 & 2 & 6 \\
\hdashline
2 & 4 & 2 & 2 & $-$ & 7 & 2 & 4 & 7 \\
2 & 4 & 4 & $-$ & 4 & 8 & 2 & 4 & 7 \\
4 & 2 & 2 & $-$ & 2 & 7 & 4 & 2 & 8 \\
4 & 2 & 4 & 4 & $-$ & 8 & 4 & 2 & 8 \\
\hline
\end{tabular}
\end{footnotesize}
\end{table}

Table \ref{tab:hh-level-re-IDs} shows an example of the identifiers for the household-level random effects in the measurement and structural models. In this example we consider the records for four individuals in three households: H1 with individuals 1 and 2, H2 with individual 3, and H3 with individual 4. There are five individual dyads for between-household ties across three household dyads. All household identifiers have the same values for individual dyads from the same household dyad, e.g. (1,3) and (2,3) from (H1,H2).

\begin{sidewaystable}
\centering
\caption{Example of identifiers for household-level random effects.  The records for three individual dyads for the between-household ties of four individuals from three households H1, H2 and H3 are shown.}
\label{tab:hh-level-re-IDs}
\begin{footnotesize}
\begin{tabular}{cccccccccccc}
\hline
 &  &  &  &  &  & \multicolumn{6}{c}{Identifiers for random effects} \\
\cline{7-12} 
Actor & Partner & Reporter & Actor hh & Partner hh & Reporter hh & & & & & & \\
$i$ & $j$ & $m$ & $h(i)$ & $h(j)$ & $h(m)$ & $r_{a,h(i)}^{(2)}$ & $r_{p,h(j)}^{(2)}$ & $r_{d|h(i)h(j)|h(m)}^{(2)}$ & $a_{h(i)}^{(2)}$ & $p_{h(j)}^{(2)}$ & $d_{h(i),h(j)}^{(2)}$\\
\hline
1 & 3 & 1 & H1 & H2 & H1 & 1 & $-$ & 1 & 1 & 2 & 1 \\
1 & 3 & 3 & H1 & H2 & H2 & $-$ & 2 & 2 & 1 & 2 & 1 \\
3 & 1 & 1 & H2 & H1 & H1 & $-$ & 1 & 1 & 2 & 1 & 2 \\
3 & 1 & 3 & H2 & H1 & H2 & 2 & $-$ & 2 & 2 & 1 & 2 \\
\hdashline
2 & 3 & 2 & H1 & H2 & H1 & 1 & $-$ & 1 & 1 & 2 & 1 \\
2 & 3 & 3 & H1 & H2 & H2 & $-$ & 2 & 2 & 1 & 2 & 1 \\
3 & 2 & 2 & H2 & H1 & H1 & $-$ & 1 & 1 & 2 & 1 & 2 \\
3 & 2 & 3 & H2 & H1 & H2 & 2 & $-$ & 2 & 2 & 1 & 2 \\
\hdashline
1 & 4 & 1 & H1 & H3 & H1 & 1 & $-$ & 3 & 1 & 3 & 3 \\
1 & 4 & 4 & H1 & H3 & H3 & $-$ & 3 & 4 & 1 & 3 & 3 \\
4 & 1 & 1 & H3 & H1 & H1 & $-$ & 1 & 3 & 3 & 1 & 4 \\
4 & 1 & 4 & H3 & H1 & H3 & 3 & $-$ & 4 & 3 & 1 & 4 \\
\hdashline
2 & 4 & 2 & H1 & H3 & H1 & 1 & $-$ & 3 & 1 & 3 & 3 \\
2 & 4 & 4 & H1 & H3 & H3 & $-$ & 3 & 4 & 1 & 3 & 3 \\
4 & 2 & 2 & H3 & H1 & H1 & $-$ & 1 & 3 & 3 & 1 & 4 \\
4 & 2 & 4 & H3 & H1 & H3 & 3 & $-$ & 4 & 3 & 1 & 4 \\
\hdashline
3 & 4 & 3 & H2 & H3 & H2 & 2 & $-$ & 5 & 2 & 3 & 5 \\
3 & 4 & 4 & H2 & H3 & H3 & $-$ & 3 & 6 & 2 & 3 & 5 \\
4 & 3 & 3 & H3 & H2 & H2 & $-$ & 2 & 5 & 3 & 2 & 6 \\
4 & 3 & 4 & H3 & H2 & H3 & 3 & $-$ & 6 & 3 & 2 & 6 \\
\hline
\end{tabular}
\end{footnotesize}
\end{sidewaystable}

\newpage
\section{Covariance structure implied by the multilevel Social Relations Model}
\label{sec:covstruc}

\subsection{Introduction}
\label{subsec:covstruc_intro}

As in the paper, we denote by $y_{i,j,m}$ a binary response coded 1 if individual $i$ (the actor) is reported to give help to individual $j$ (the partner) by reporter $m \in \{i,j\}$. Denote by $h(i)$ and $h(j)$ the households of individuals $i$ and $j$, referred to as the actor and partner household for the individual dyad $(i,j)$. The multilevel SRM is specified in terms of $y_{i,j,m}^*$ an underlying continuous latent response such that $y_{i,j,m}=\mbox{I}(y_{i,j,m}^*>0)$ where $\mbox{I}(\cdot)$ is the indicator function.

The multilevel SRM comprises a measurement model, eq. (\ref{eq:model_measurement}), and a structural model, eq. (\ref{eq:model_structural}), where the assumptions and interpretation of the random effects are given in the paper.
\begin{eqnarray}
\label{eq:model_measurement}
y_{i,j,m}^* & = & z_{i,j} + \mbox{I}(m=i) (\beta_a + r_{ai}^{(1)} + r_{a,h(i)}^{(2)}) + \mbox{I}(m=j) (\beta_p + r_{pj}^{(1)} + r_{p,h(j)}^{(2)}) \\ \nonumber
& + & r_{d |i,j| m}^{(1)} + r_{d |h(i),h(j)| h(m)}^{(2)} + \epsilon_{i,j,m},
\end{eqnarray}
\begin{equation}
\label{eq:model_structural}
z_{i,j} = a_i^{(1)} + p_j^{(1)} + d_{i,j}^{(1)} + a_{h(i)}^{(2)} + p_{h(j)}^{(2)} + d_{h(i),h(j)}^{(2)}.
\end{equation}

Under this model, the variances of $y_{i,j,m}^*$, depend on whether the reporter $m$ is the actor ($m=i$) or partner ($m=j$) as follows:
\begin{eqnarray}
\label{eq:V_ra.hh}
V_{ra} & = & \sigma_{a(1)}^2 + \sigma_{p(1)}^2 + \sigma_{d(1)}^2 + \sigma_{a(2)}^2 + \sigma_{p(2)}^2 +  \sigma_{d(2)}^2\\ \nonumber
& + & \sigma_{ra(1)}^2 + \sigma_{rd(1)}^2 + \sigma_{ra(2)}^2 + \sigma_{rd(2)}^2 + 1 \nonumber
\end{eqnarray}
\begin{eqnarray}
\label{eq:V_rp.hh}
V_{rp} & = & \sigma_{a(1)}^2 + \sigma_{p(1)}^2 + \sigma_{d(1)}^2 + \sigma_{a(2)}^2 + \sigma_{p(2)}^2 +  \sigma_{d(2)}^2\\ \nonumber
& + & \sigma_{rp(1)}^2 + \sigma_{rd(1)}^2 + \sigma_{rp(2)}^2 + \sigma_{rd(2)}^2 + 1. \nonumber
\end{eqnarray}

\noindent The model implies that the covariance between a pair of responses $(y_{i,j,m},y_{i^\prime,j^\prime,m^\prime})$ is non-zero when the pair share individuals or household members.  These non-zero covariances are described in Table \ref{tab:nonzero_cov} where we distinguish `within-dyad' pairs, involving the same individuals $i$ and $j$, and `cross-dyad' pairs that have at most one individual in common.  The next sections give the form of the within-dyad and cross-dyad covariances in terms of the model parameters. For all other pairs of responses, the covariance is assumed to be zero, i.e.
\begin{equation*}
\mbox{cov}(y_{i,j,m}, y_{i^\prime,j^\prime,m^\prime}) = 0 \, \mbox{for} \, i \neq i^\prime, j \neq j^\prime, m \in \{i,j \}, m^\prime \in \{i^\prime,j^\prime \} \, \mbox{and} \, h(k) \neq h(k^\prime), \, k \neq k^\prime \in \{i,j,i^\prime,j^\prime\}.
\end{equation*}

\begin{table} [!htbp]
\caption{Nonzero covariances under the model given by eq. (\ref{eq:model_measurement}) and (\ref{eq:model_structural}) where $i$ indexes the actor, $j$ the partner and $m$ the reporter, and $h(i)$, $h(j)$ and $h(m)$ denote their corresponding households.  Expressions for the covariances in terms of the model parameters are given in Sections \ref{subsec:covstruc_wd}, \ref{subsec:covstruc_cd_ind} and \ref{subsec:covstruc_cd_hh}.}
\label{tab:nonzero_cov}
\begin{footnotesize}
\begin{tabular}{lll}
\hline
Covariance & Description & Label \\
$\mbox{cov}(y_{i,j,m}^*, y_{i^\prime,j^\prime,m^\prime}^*)$ & & \\[3mm]
\hline
\multicolumn{3}{l}{Within-dyad} \\
$i^\prime=i, \, j^\prime=j, \, m=i, \, m^\prime=j$  & Same directed dyad, different reporters & $c_1$ \\
$i^\prime=j, \, j^\prime=i, \, m=m^\prime=i$  & Same dyad, same reporter & $c_2$ \\
$i^\prime=j, \, j^\prime=i, \, m=j, \, m^\prime=i$ & Same dyad, different reporters, partner reporting on each & $c_3$ \\
$i^\prime=j, \, j^\prime=i, \, m=i, \, m^\prime=j$ & Same dyad, different reporters, actor reporting on each & $c_3$ \\[3mm]
\multicolumn{3}{l}{Cross-dyad (dyads share one individual)$^\dagger$} \\
$i^\prime=i, \, m=m^\prime=i$ & Same actor, diff. partners, same reporter (actor) & $c_4$ \\
$i^\prime=i, \, m=j, \, m^\prime=j^\prime$ & Same actor, diff. partners, diff. reporters (partners) & $c_5$ \\
$i^\prime=i, \, m=i, \, m^\prime=j^\prime$ & Same actor, diff. partners, diff. reporters (actor/partner) & $c_6$ \\
$j^\prime=j, \, m= m^\prime=j$ & Same partner, diff. actors, same reporter (partner) & $c_7$ \\
$j^\prime=j, \, m=i, \, m^\prime=i^\prime$  & Same partner, diff. actors, diff. reporters (actors) & $c_8$ \\
$j^\prime=j, \, m=i, \, m^\prime=j$ & Same partner, diff. actors, diff. reporters (actor/partner) & $c_9$ \\
$j^\prime=i, \, m=m^\prime=i$ & Same person is actor in one dyad, partner in other, same reporter (actor/partner) & $c_{10}$ \\
$j^\prime=i, \, m=i, \, m^\prime=i^\prime$ & Same person is actor in one dyad, partner in other, diff. reporters (actors) & $c_{11}$ \\
$j^\prime=i, \, m=j, \, m^\prime=i$ & Same person is actor in one dyad, partner in other, diff. reporters (partners) & $c_{12}$ \\
$j^\prime=i, \, m=j, \, m^\prime=i^\prime$ & Same person is actor in one dyad, partner in other, diff. reporters (actor/partner) & $c_{13}$ \\[3mm]
\multicolumn{3}{l}{Cross-dyad (dyads share household members, but not an individual)$^\ddagger$} \\
$h(i)=h(i^\prime), \, h(j)=h(j^\prime)$ & Actors in same hh, partners in same hh & $c_{14}$ \\
$h(i)=h(i^\prime)$ & Actors in same hh, partners in diff. hh & $c_{15}$ \\
$h(j)=h(j^\prime)$ & Partners in same hh, actors in diff. hh & $c_{16}$ \\
$h(i)=h(j^\prime)$ or $h(j)=h(i^\prime)$ & Actor in one dyad in same hh as partner in other dyad & $c_{17}$ \\
\hline
\multicolumn{3}{p{0.99\textwidth}}{\vspace{1.5mm}$^\dagger$\footnotesize{Cross-dyad covariances where dyads share an individual additionally depend on whether dyads also share individuals from the same household}; $^\ddagger$\footnotesize{Cross-dyad covariances where dyads share household members additionally depend on whether reporters are actors, partners or a combination. 
} }
\end{tabular}
\end{footnotesize}
\end{table}

\subsection{Within-dyad covariances}
\label{subsec:covstruc_wd}

The measurement model (\ref{eq:model_measurement}) takes the following form for the four latent responses in a given dyad:
\begin{gather}
  y_{i,j,i}^* = z_{i,j} + \beta_a + r_{ai}^{(1)} + r_{d |i,j| i}^{(1)} + r_{a,h(i)}^{(2)} + r_{d |h(i),h(j)| h(i)}^{(2)} + \epsilon_{i,j,i} \label{eq:y_iji} \\
  y_{i,j,j}^* = z_{i,j} + \beta_p + r_{pj}^{(1)} + r_{d |i,j| j}^{(1)} + r_{p,h(j)}^{(2)} + r_{d |h(i),h(j)| h(j)}^{(2)} + \epsilon_{i,j,j} \label{eq:y_ijj} \\
  y_{j,i,i}^* = z_{j,i} + \beta_p + r_{pi}^{(1)} + r_{d |i,j| i}^{(1)} + r_{p,h(i)}^{(2)} + r_{d |h(i),h(j)| h(i)}^{(2)}+ \epsilon_{j,i,i} \label{eq:y_jii} \\
  y_{j,i,j}^* = z_{j,i} + \beta_a + r_{aj}^{(1)} + r_{d |i,j| j}^{(1)} + r_{a,h(j)}^{(2)} + r_{d |h(i),h(j)| h(j)}^{(2)}+ \epsilon_{j,i,j}. \label{eq:y_jij}
\end{gather}

Under the model assumptions, the form of the implied within-dyad covariance matrix is shown in Table \ref{tab:wd_cov}, using the labels $(c_1, c_2, c_3)$ given in Table \ref{tab:nonzero_cov}. The three distinct covariances $(c_1, c_2, c_3)$ in Table \ref{tab:wd_cov} are given by:
\begin{align*}
  c_1 & = \sigma_{a(1)}^2 + \sigma_{p(1)}^2 + \sigma_{d(1)}^2 + \sigma_{a(2)}^2 + \sigma_{p(2)}^2 + \sigma_{d(2)}^2, \\[2ex]
  c_2 & = 2 \sigma_{ap(1)} + \sigma_{dd(1)} + 2 \sigma_{ap(2)} + \sigma_{dd(2)} + \sigma_{rap(1)} + \sigma_{rd(1)}^2 +
   \sigma_{rap(2)} +  \sigma_{rd(2)}^2,  \\[2ex]
  c_3 & = 2 \sigma_{ap(1)} + \sigma_{dd(1)} + 2 \sigma_{ap(2)} + \sigma_{dd(2)}.
\end{align*}
To illustrate the derivation of the covariances, consider $c_1$ for which we use the random parts of eq. (\ref{eq:y_iji}) and (\ref{eq:y_ijj}):
\begin{align*}
  c_1 = & \mbox{cov}(y_{i,j,i}^*, y_{i,j,j}^*) \\
    = & \mbox{cov}(z_{i,j} + r_{ai}^{(1)} + r_{d |i,j| i}^{(1)} + r_{a,h(i)}^{(2)} + r_{d |h(i),h(j)| h(i)}^{(2)} + \epsilon_{i,j,i},\\
     & ~~~~~  z_{i,j} + r_{pj}^{(1)} + r_{d |i,j| j}^{(1)} + r_{p,h(j)}^{(2)} + r_{d |h(i),h(j)| h(j)}^{(2)} + \epsilon_{i,j,j}) \\
  = & \mbox{var}(z_{i,j})  \\
   = & \sigma_{a(1)}^2 + \sigma_{p(1)}^2 + \sigma_{d(1)}^2 + \sigma_{a(2)}^2 + \sigma_{p(2)}^2 + \sigma_{d(2)}^2
\end{align*}
\noindent as all covariances equal zero under the model assumptions.

\begin{table}[!htbp]
\caption{Structure of the within-dyad covariance matrix for the model given by eq. (\ref{eq:model_measurement}) and (\ref{eq:model_structural})}
\label{tab:wd_cov}
\begin{center}
\begin{tabular}{lcccc}
\hline
                & $y_{i,j,i}^*$     & $y_{i,j,j}^*$     & $y_{j,i,i}^*$     & $y_{j,i,i}^*$ \\[2ex]
\hline
$y_{i,j,i}^*$   & $V_{ra}$          &                   &                   &       \\[2ex]
$y_{i,j,j}^*$   & $c_1$             &  $V_{rp}$         &                   &       \\[2ex]
$y_{j,i,i}^*$   & $c_2$             &  $c_3$            & $V_{rp}$          &       \\[2ex]
$y_{j,i,j}^*$   & $c_3$             &  $c_2$            & $c_1$             & $V_{ra}$ \\
\hline
\end{tabular}
\end{center}
\end{table}

\subsection{Cross-dyad covariances when dyads share an individual}
\label{subsec:covstruc_cd_ind}

Non-zero cross-dyad covariances arise when the dyads $(i,j)$ and $(i^\prime, j^\prime)$ share one individual.  The form of these covariances in terms of the model parameters is given below, again using the labels from Table \ref{tab:nonzero_cov} and assuming $i \neq j $ and $i^\prime \neq j^\prime$. In each case, there is an additional component according to whether a pair of dyads share individuals from the same household (assuming that individuals in the same dyad are from different households).  For example, the covariance between $y_{i,j,i}^*$ and $y_{i,j^\prime,i}^*$ ($c_4$), has three additional variance components when the partners $j$ and $j^\prime$ are from the same household: $\sigma_{p(2)}^2$ (from the household-level partner effect), $\sigma_{d(2)}^2$ (from the household-level dyad effect as individual dyads $(i,j)$ and $(i,j^\prime)$ are from the same household dyad) and $\sigma_{rd(2)}^2$ (from the household reporter dyad effect as $i$ reports on the same household dyad).

\begin{align*}
c_4 & = \mbox{cov}(y_{i,j,i}^*, y_{i,j^\prime,i}^*) = \sigma_{a(1)}^2 + \sigma_{a(2)}^2 + \sigma_{ra(1)}^2 + \sigma_{ra(2)}^2 +
\mbox{I}\{h(j)=h(j^\prime)\} (\sigma_{p(2)}^2 + \sigma_{d(2)}^2 + \sigma_{rd(2)}^2)\\
c_5 & = \mbox{cov}(y_{i,j,j}^*, y_{i,j^\prime,j^\prime}^*) = \sigma_{a(1)}^2 + \sigma_{a(2)}^2 +
\mbox{I}\{h(j)=h(j^\prime)\} (\sigma_{p(2)}^2 + \sigma_{d(2)}^2 + \sigma_{rp(2)}^2 + \sigma_{rd(2)}^2)  \\
c_6 & = \mbox{cov}(y_{i,j,i}^*, y_{i,j^\prime,j^\prime}^*) = \sigma_{a(1)}^2 + \sigma_{a(2)}^2 +
\mbox{I}\{h(j)=h(j^\prime)\} (\sigma_{p(2)}^2 + \sigma_{d(2)}^2 )  \\
c_7 & = \mbox{cov}(y_{i,j,j}^*, y_{i^\prime,j,j}^*) = \sigma_{p(1)}^2 + \sigma_{p(2)}^2 + \sigma_{rp(1)}^2 + \sigma_{rp(2)}^2 +
\mbox{I}\{h(i)=h(i^\prime)\} (\sigma_{a(2)}^2 + \sigma_{d(2)}^2 + \sigma_{rd(2)}^2)  \\
c_8 & = \mbox{cov}(y_{i,j,i}^*, y_{i^\prime,j,i^\prime}^*) = \sigma_{p(1)}^2 + \sigma_{p(2)}^2 +
\mbox{I}\{h(i)=h(i^\prime)\} (\sigma_{a(2)}^2 + \sigma_{d(2)}^2 + \sigma_{ra(2)}^2 + \sigma_{rd(2)}^2) \\
c_9 & = \mbox{cov}(y_{i,j,i}^*, y_{i^\prime,j,j}^*) = \sigma_{p(1)}^2 + \sigma_{p(2)}^2 +
\mbox{I}\{h(i)=h(i^\prime)\} (\sigma_{a(2)}^2 + \sigma_{d(2)}^2 ) \\
c_{10} & = \mbox{cov}(y_{i,j,i}^*, y_{i^\prime,i,i}^*) = \sigma_{ap(1)} + \sigma_{ap(2)} + \sigma_{rap(1)} + \sigma_{rap(2)} +
\mbox{I}\{h(j)=h(i^\prime)\} (\sigma_{ap(2)} + \sigma_{dd(2)} + \sigma_{rd(2)}^2 ) \\
c_{11} & = \mbox{cov}(y_{i,j,i}^*, y_{i^\prime,i,i^\prime}^*) = \sigma_{ap(1)} + \sigma_{ap(2)} +
\mbox{I}\{h(j)=h(i^\prime)\} (\sigma_{ap(2)} + \sigma_{dd(2)} ) \\
c_{12} & = \mbox{cov}(y_{i,j,j}^*, y_{i^\prime,i,i}^*) = \sigma_{ap(1)} + \sigma_{ap(2)} +
\mbox{I}\{h(j)=h(i^\prime)\} (\sigma_{ap(2)} + \sigma_{dd(2)} ) \\
c_{13} & = \mbox{cov}(y_{i,j,j}^*, y_{i^\prime,i,i^\prime}^*) = \sigma_{ap(1)} + \sigma_{ap(2)} +
\mbox{I}\{h(j)=h(i^\prime)\} (\sigma_{ap(2)} + \sigma_{dd(2)} + \sigma_{rap(2)} +  \sigma_{rd(2)}^2 ) \\
\end{align*}

\subsection{Cross-dyad covariances when dyads share household members}
\label{subsec:covstruc_cd_hh}

Finally, non-zero cross-dyad covariances arise when the two dyads share members of the same household.  The form of the covariances for four general cases (described in Table \ref{tab:nonzero_cov}) are given below, where $h(\cdot)$ indicates the household membership of an individual. In each case, any pair of individuals other than those specified are assumed to be in different households.  For example, in $c_{15}$ the actors $i$ and $i^\prime$ are in the same household, but the partners $j$ and $j^\prime$ are in different households.  In this case, the covariance equals the variance of the household-level actor effect, with an additional contribution from the household-level actor reporter effect when both reports are from the actors (in which case $m=i, m^\prime=i^\prime$ which implies $h(m)=h(m^\prime)$).

\begin{align*}
c_{14} & = \mbox{cov}(y_{i,j,m}^*, y_{i^\prime,j^\prime,m^\prime}^*) \; \mbox{where} \; h(i)=h(i^\prime), \, h(j)=h(j^\prime)\\
& = \sigma_{a(2)}^2 + \sigma_{p(2)}^2 + \sigma_{d(2)}^2 + \mbox{I}\{h(m)=h(m^\prime)\} \, [\sigma_{rd(2)}^2 + 
\mbox{I}\{m=i,\, m^\prime=i^\prime\} \, \sigma_{ra(2)}^2 + \mbox{I}\{m=j,\, m^\prime=j^\prime\} \, \sigma_{rp(2)}^2 \\
& + \mbox{I}\{m=j,\, m^\prime=i^\prime\} \, \sigma_{rap(2)}
+ \mbox{I}\{m=i,\, m^\prime=j^\prime\} \, \sigma_{rap(2)}]\\[3mm]
c_{15} & = \mbox{cov}(y_{i,j,m}^*, y_{i^\prime,j^\prime,m^\prime}^*) \; \mbox{where} \; h(i)=h(i^\prime)\\
& = \sigma_{a(2)}^2 + \mbox{I}\{m=i,\, m^\prime=i^\prime\} \, \sigma_{ra(2)}^2 \\[3mm]
c_{16} & = \mbox{cov}(y_{i,j,m}^*, y_{i^\prime,j^\prime,m^\prime}^*) \; \mbox{where} \; h(j)=h(j^\prime)\\
& = \sigma_{p(2)}^2 + \mbox{I}\{m=j,\, m^\prime=j^\prime\} \, \sigma_{rp(2)}^2 \\[3mm]
c_{17} & = \mbox{cov}(y_{i,j,m}^*, y_{i^\prime,j^\prime,m^\prime}^*) \; \mbox{where} \; h(i)=h(j^\prime)\\
& = \sigma_{ap(2)} + \mbox{I}\{m=i,\, m^\prime=j^\prime\} \, \sigma_{rap(2)}\\
\end{align*}

\subsection{Estimated correlations for social support in Nicaragua}
\label{subsec:covstruc_results}

Estimates of the within-dyad and selected cross-dyad correlations from the multilevel SRM for social support are given in Table \ref{tab:nonzero_cov_nic}.  The posterior means and 95\% credible intervals for the correlations are shown, based on the relevant covariances $(c_1,\ldots, c_{17})$ defined in Sections \ref{subsec:covstruc_wd}, \ref{subsec:covstruc_cd_ind} and \ref{subsec:covstruc_cd_hh} and the appropriate variances given by eq. (\ref{eq:V_ra.hh}) and (\ref{eq:V_rp.hh}).  For example, the within-dyad correlation based on the covariance $c_1$ is 
\begin{equation*}
  \mbox{cor}(y_{i,j,i}^*, y_{i,j,j}^*) =  \frac{\mbox{cov}(y_{i,j,i}^*, y_{i,j,j}^*)}{\sqrt{V_{ra}V_{rp}}} 
\end{equation*}
where the denominator is the square root of the product of the actor and partner variances given by eq. (\ref{eq:V_ra.hh}) and (\ref{eq:V_rp.hh}) because the first response is an actor report while the second is a partner report.  When the correlation is between two responses from the same reporter, the denominator is simply the appropriate variance.  For example, the within-dyad correlation based on the covariance $c_2$ is
\begin{equation*}
  \mbox{cor}(y_{i,j,i}^*, y_{j,i,j}^*) =  \frac{\mbox{cov}(y_{i,j,i}^*, y_{j,i,j}^*)}{V_{ra}} 
\end{equation*}
because both responses are actor reports.

The estimates of the within-dyad correlations from the model are broadly similar to the sample tetrachoric correlations presented in Section 5.2 of the paper, with the exception of the estimate of the correlation between partner reports of support given and received ($\mbox{cor}(y_{i,j,j}^*, y_{j,i,i}^*)$, based on the covariance $c_3$) which is estimated as 0.566 from the model compared with a sample tetrachoric correlation of 0.429. Under the model, $\mbox{cov}(y_{i,j,j}^*, y_{j,i,i}^*) = \mbox{cov}(y_{i,j,i}^*, y_{j,i,j}^*)$ so the difference in the \emph{correlations} between support given and received for actor and partner reports comes entirely from differences in the actor and partner variances which, in this case, is not sufficient to capture the difference in the correlations. 

Turning to the estimates of the cross-dyad correlations, there is a moderate correlation of 0.322 between an individual $i$'s reports of giving help to two different individuals $j$ and $j^\prime$ (based on covariance $c_4$), which reflects a combination of actor and reporter effects (when the reporter is the actor/giver).  This correlation increases to 0.614 when the partners/recipients $j$ and $j^\prime$ are in the same household. When reporter effects are absent (i.e. when reports are from two different individuals), these correlations decrease to 0.114 and 0.391 respectively.  A similar pattern is found when dyads have the partner rather than actor in common (based on covariances $c_7$ and $c_8$, results not shown).

Finally, we consider cross-dyad correlations when the dyads share household members, but have no individual in common.  There is a correlation of 0.277 between reports of individuals in the same household giving help to individuals who are also in the same household (but different to the actors'/givers' household) (based on covariance $c_{14}$). This correlation increases to 0.316 when the pair of reports are from actors in the same household. When only the actors are in the same household, the correlation between reports of giving to individuals in different households is negligible ($c_{15}$). Similar patterns are found when the partners rather than actors are in the same household (based on covariances $c_{16}$ and $c_{17}$, results not shown).

\begin{table} [!htbp]
\caption{Estimates of selected nonzero covariances for social support outcomes in Nicaragua}
\label{tab:nonzero_cov_nic}
\centering
\begin{footnotesize}
\begin{tabular}{lcc}
\hline
Label: Description & Mean & 95\% CI \\
\hline
\multicolumn{3}{l}{Within-dyad} \\
$c_{1}$: Same directed dyad, different reporters & 0.584 & (0.533, 0.632) \\
$c_{2}$: Same dyad, same reporter & 0.845 & (0.818, 0.867) \\
$c_{3}$: Same dyad, different reporters, partner reporting on each & 0.566 & (0.513, 0.617)  \\
$c_{3}$: Same dyad, different reporters, actor reporting on each & 0.525 & (0.470, 0.579)  \\ 
\addlinespace[1ex]
\multicolumn{3}{l}{Cross-dyad (dyads share an individual)} \\
$c_{4}$: Same actor, diff. partners, same reporter (actor) & 0.322 & (0.275, 0.372)  \\
$c_{4}$: $+$ partners in same hh & 0.614 & (0.577, 0.655)  \\
$c_{5}$: Same actor, diff. partners, diff. reporters (partners) & 0.114 & (0.087, 0.148)  \\
$c_{5}$: $+$ partners in same hh & 0.391 & (0.344, 0.445)  \\
$c_{7}$: Same partner, diff. actors, same reporter (partner) & 0.364 & (0.313, 0.419)   \\
$c_{7}$: $+$ actors in same hh & 0.635 & (0.595, 0.676)  \\
$c_{8}$: Same partner, diff. actors, diff. reporters (actors) & 0.116 & (0.087, 0.149)   \\
$c_{8}$: $+$ actors in same hh & 0.415 & (0.368, 0.468)  \\
\addlinespace[1ex]
\multicolumn{3}{l}{Cross-dyad (dyads share household members, but not an individual)} \\
$c_{14}$: Actors in same hh, partners in same hh, reporters (actor/partner) in diff. hh & 0.277 & (0.226, 0.339) \\
$c_{14}$: Actors in same hh, partners in same hh, reporters (actors) in same hh & 0.316 & (0.261, 0.383) \\
$c_{15}$: Actors in same hh, partners in diff. hh, reporters (partners) in diff. hh & 0.015 & (0.0001, 0.049) \\
\hline
\end{tabular}
\end{footnotesize}
\end{table}

\section{Model identification and estimability}

\subsection{Identification of random effect parameters}

Following the discussion of identification of the dyadic IRT model in \cite{gin.etal2020}, the model is identified if the $p$ model parameters for the covariance structure of the latent responses $y_{i,j,m}^*$ can each be uniquely identified by the $r$ distinct model-implied variances and covariances of $y_{i,j,m}^*$ (referred to by \citeauthor{gin.etal2020} as the ``composite latent variable'').  A necessary condition for identification is that $p \leq r$. The model has $p=18$ random effect variance and covariance parameters (8 in the measurement model and 10 in the structural model) which describe the covariance structure of $y_{i,j,m}^*$. The covariance structure of $y_{i,j,m}^*$ implied by the multilevel SRM are linear combinations of the $p$ random effect parameters, as set out in Section \ref{sec:covstruc}.  There are two variances, $V_{ra}$ and $V_{rp}$, depending on whether reporter $m$ is the actor or partner in the dyad, and 33 covariances $\mbox{cov}(y_{i,j,m}^*, y_{i^\prime,j^\prime,m^\prime}^*)$ given by the expressions $c_1, \ldots, c_{17}$. Expressions $c_1,c_2,c_3$ define the within-dyad covariances, and $c_4, \ldots, c_{17}$ define the cross-dyad covariances which depend on whether the dyads and reporters $(i,j,m)$ and $(i^\prime,j^\prime,m^\prime)$ are formed of individuals from the same household. The total of 33 covariances comprise 3 from $c_1,c_2,c_3$, 20 from $c_4,\ldots,c_{13}$, 4 from $c_{14}$ and 6 from $c_{15},c_{16},c_{17}$.  However, not all covariance expressions are unique functions of the $p$ parameters, so $r<35$.  For example, $c_{11}=c_{12}$ and, when $j$ and $i^\prime$ are in different households, $c_{11}=c_{12}=c_{13}$, thus reducing $r$ by 3.

We begin by showing that the SRM without household effects is identified. As this model includes only individual-level random effects, the `(1)' superscripts are omitted to obtain
\begin{equation}
y_{i,j,m}^* = z_{i,j} + I(m=i) (\beta_1 + r_{ai}) + I(m=j) (\beta_2 + r_{pj}) + r_{d |ij| m} + \epsilon_{i,j,m} \quad m \in \{ i,j \}\\
\end{equation}
\begin{equation}
z_{i,j} = a_i + p_j + d_{i,j}
\end{equation}
where the model assumptions are given in Sections 3.1 and 3.2 of the paper. The model has 9 random effect parameters: $(\sigma_{ra}^2, \sigma_{rp}^2, \sigma_{rap}, \sigma_{rd}^2)$ in the measurement model and $(\sigma_{a}^2, \sigma_{p}^2, \sigma_{ap}, \sigma_d^2, \sigma_{dd})$ in the structural model.

When household effects are omitted, the expressions for the within-dyad covariances in Section \ref{subsec:covstruc_wd} reduce to:
\begin{align*}
  c_1 & = \mbox{cov}(y_{i,j,i}^*, y_{i,j,j}^*) = \sigma_{a}^2 + \sigma_{p}^2 + \sigma_{d}^2,  \\[2ex]
  c_2 & = \mbox{cov}(y_{i,j,i}^*, y_{j,i,i}^*) = 2 \sigma_{ap} + \sigma_{dd} + \sigma_{rap} + \sigma_{rd}^2,  \\[2ex]
  c_3 & = \mbox{cov}(y_{i,j,i}^*, y_{j,i,j}^*) = 2 \sigma_{ap} + \sigma_{dd}.
\end{align*}

The 10 expressions for the cross-dyad covariances when dyads share an individual in Section \ref{subsec:covstruc_cd_ind} reduce to the following 6 distinct expressions:
\begin{align*}
c_4 & = \mbox{cov}(y_{i,j,i}^*, y_{i,j^\prime,i}^*) = \sigma_{a}^2 + \sigma_{ra}^2,\\
c_5 =c_6 & = \mbox{cov}(y_{i,j,j}^*, y_{i,j^\prime,j^\prime}^*) = \sigma_{a}^2, \\
c_7 & = \mbox{cov}(y_{i,j,j}^*, y_{i^\prime,j,j}^*) = \sigma_{p}^2 + \sigma_{rp}^2,   \\
c_8=c_9 & = \mbox{cov}(y_{i,j,i}^*, y_{i^\prime,j,i^\prime}^*) = \sigma_{p}^2, \\
c_{10} & = \mbox{cov}(y_{i,j,i}^*, y_{i^\prime,i,i}^*) = \sigma_{ap} + \sigma_{rap},  \\
c_{11}=c_{12}=c_{13} & = \mbox{cov}(y_{i,j,i}^*, y_{i^\prime,i,i^\prime}^*) = \sigma_{ap}.  \\
\end{align*}

Finally, the 4 cross-dyad covariances $c_{14},\ldots,c_{17}$ for dyads that share household members, but not an individual, simplify to zero when household effects are excluded.

It is straightforward to show that the 9 equations for the 9 distinct covariances $(c_4, c_5, c_7, c_8, c_{10},c_{11})$ can be solved for the 9 random effect parameters, without requiring the equations for the variances $V_{ra}$ and $V_{rp}$.

\begin{align*}
  \sigma_a^2 & = c_5 \\
  \sigma_p^2 & = c_8 \\
  \sigma_{ap} & = c_{11} \\
  \sigma_{ra}^2 & = c_4 - c_5 \\
  \sigma_{rp}^2 & = c_7 - c_8 \\
  \sigma_{rap} & = c_{10} - c_{11} \\
  \sigma_d^2 & = c_1 - c_5 - c_8 \\
  \sigma_{dd} & = c_3 - 2 c_{11} \\
  \sigma_{rd}^2 & = c_2 - 2 c_{11} - (c_3 - 2 c_{11}) - (c_{10} - c_{11}).
\end{align*}

Returning to the full multilevel SRM with household effects, it is more difficult to show identification using this approach due to the large number of equations defined by $c_1, \ldots, c_{17}$.  More generally, we can write the equations for the model-implied variances and covariances of $y_{i,j,m}^*$ as $\mathbf{A} \mathbf{\theta} = \mathbf{c}$ where $\mathbf{\theta}$ is the $p$-vector of random effect parameters, $\mathbf{c}$ is the $q$-vector of model-implied variances and covariances and $\mathbf{A}$ is a $q \times p$ matrix ($q \leq r$ where $r$ is the number of distinct model-implied variances and covariances). If the rank (number of linearly independent rows) of $\mathbf{A}$ is $p$, each element of $\mathbf{\theta}$ can be expressed as a unique linear combination of the elements of $\mathbf{c}$.  For the multilevel SRM, we find that $\mbox{rank}(\mathbf{A})=p=18$, so we conclude that the model parameters are identified.

\subsection{Estimability}

Although the parameters of the multilevel SRM model can be uniquely identified in theory, it is important to assess whether the model is estimable for a given data structure, i.e. whether the quantities of interest can be reliably estimated from the available data. To check this for our application, we carried out a small-scale simulation study where the model was fitted to data generated to have the same multilevel structure as the Nicaraguan data. The data generation model (DGM) is the multilevel SRM with parameter values set to be similar to the posterior means obtained from fitting the model to the real data.  The model was fitted to 50 simulated datasets where different random effect draws were taken to obtain a different set of binary responses $y_{i,j,m}$ for each dataset. For each replication, the posterior means and standard deviations of the parameters are based on a single chain of 5000 MCMC iterations, with a warm-up sample of 3000 and no thinning. These settings were found to be sufficient for convergence. The number of replications is relatively small due to long estimation times, so the results should be taken as indicative.

Of primary substantive interest are the variance partitioning coefficients (VPCs) and random effect correlations from the measurement and structural models (as shown in Tables 2 and 3 of the paper). The ``true'' values of the VPCs were derived from the parameter values assumed in the DGM. The simulation results for the measurement and structural models are shown in Tables \ref{tab:sim_results_meas} and \ref{tab:sim_results_struct} respectively. Values in the `mean' and `mean SE' columns are the means of the posterior means and standard deviations of the VPCs and correlations across replications.  Values in the `SD' column are the empirical standard errors, calculated as the standard deviation of the posterior means. The Monte Carlo errors (MCEs) for the mean and the empirical SE are given in `MCE(Mean)' and `MCE(SD)'.  MCE(Mean) is calculated as $\mbox{SD}/\sqrt{R}$ where $R=50$ is the number of replications.  MCE(SD) was calculated using the \texttt{mcstatsim} R package \citep{mcstatsim} which takes account of the kurtosis of the sampling distribution; this too depends inversely on $R$.

For both the measurement and structural models, the mean estimates of most VPCs and correlations are within 2 MCEs of the true values and the mean SEs are within 2 MCEs of the empirical SEs.  This suggests that we are able to recover the true values with accurate SEs for most parameters when the model assumptions hold. There are two notable exemptions to this trend: the household-level correlations between the actor and partner reporter effects in the measurement model ($\rho_{rap(2)}$, Table \ref{tab:sim_results_meas}) and between the actor and partner effects in the structural model ($\rho_{ap(2)}$, Table \ref{tab:sim_results_struct}). There is a substantial downward bias in both parameter estimates and upward bias in the SE. The bias is likely explained by the relatively small number of households ($n=32$); although the variances of household-level random effects are based on the same number of observations, correlation estimates are more unstable because they depend on the covariance and variances of the household actor and partner effects which are all imprecisely estimated. It is notable that there is no such bias in the estimates and SEs of random effect correlations based on a larger number of observations (e.g. the individual actor and partner effects for 108 individuals). In light of these results, the household-level correlations from the analysis of the real data are interpreted with caution in the discussion in Section 5.4 of the paper. We also note that the problem can be avoided by replacing the bivariate actor- and partner-specific household effects by univariate overall household effects.

\begin{table}[!htbp]
\centering
\caption{Simulation results from fitting multilevel SRM to 50 datasets generated from the same model: Measurement model. The generated data have the same multilevel structure as the Nicaraguan data. VPCs are the percentage of the total variance in $y_{i,j,m}^*$ explained by each component of variation for actor ($m=i$) and partner reports ($m=j$).}
\label{tab:sim_results_meas}
\begin{footnotesize}
\begin{tabularx}{\textwidth}{l *{6}{R}}
\hline
Parameter & True & Mean & MCE(Mean) & Mean SE & SD & MCE(SD) \\
\hline
Actor report intercept $\beta_a$ & $-$2.10 & $-$2.15 & 0.06 & 0.34 & 0.39 & 0.18 \\
Partner-actor report difference $\beta_p - \beta_a$ & $-$0.50 & $-$0.51 & 0.02 & 0.16 & 0.14 & 0.06 \\
& & & & & & \\
\multicolumn{7}{l}{\emph{VPC as \% of total variance for actor reports, $\mbox{var}(y_{i,j,i}^*)$}} \\
Individual: Actor $r_{ai}^{(1)}$                & 19.38 & 18.98 & 0.35 & 2.59 & 2.49 & 1.06 \\
Individual: Dyad $r_{d|i,j|}^{(1)}$             &  9.30 &  9.79 & 0.14 & 1.01 & 0.98 & 0.40 \\
Household: Actor $r_{a,h(i)}^{(2)}$             &  0.78 &  1.05 & 0.09 & 1.10 & 0.65 & 0.30 \\
Household: Dyad $r_{d|h(i),h(j)|h(m)}^{(2)}$    &  2.33 &  2.36 & 0.06 & 0.50 & 0.45 & 0.18 \\
& & & & & & \\
\multicolumn{7}{l}{\emph{VPC as \% of total variance for partner reports, $\mbox{var}(y_{i,j,j}^*)$}} \\
Individual: Partner $r_{pj}^{(1)}$              & 25.00 & 24.73 & 0.45 & 3.12 & 3.16 & 1.45 \\
Individual: Dyad $r_{d|i,j|}^{(1)}$             &  8.57 &  9.04 & 0.13 & 0.96 & 0.89 & 0.35 \\
Household: Partner $r_{p,h(j)}^{(2)}$           &  1.43 &  1.37 & 0.13 & 1.38 & 0.91 & 0.44 \\
Household: Dyad $r_{d|h(i),h(j)|h(m)}^{(2)}$    &  2.14 &  2.18 & 0.06 & 0.46 & 0.41 & 0.16 \\
& & & & & & \\
\multicolumn{7}{l}{\emph{Actor-partner report correlations}}  \\
Individual: $\rho_{rap(1)}$                     &  0.80 &  0.77 & 0.01 & 0.05 & 0.06 & 0.03 \\
Household: $\rho_{rap(2)}$                      &  0.20 &  0.09 & 0.01 & 0.44 & 0.09 & 0.04 \\
\hline
\end{tabularx}
\end{footnotesize}
\end{table}

\begin{table}[!htbp]
\centering
\caption{Simulation results from fitting multilevel SRM to 50 datasets generated from the same model: Structural model. The generated data have the same multilevel structure as the Nicaraguan data. VPCs are the percentage of the total variance in $z_{i,j}$ explained by each component of variation.}
\label{tab:sim_results_struct}
\begin{footnotesize}
\begin{tabularx}{\textwidth}{l *{6}{R}}
\hline
Parameter & True & Mean & MCE(Mean) & Mean SE & SD & MCE(SD) \\
\hline
\multicolumn{7}{l}{\emph{Variance partitioning coefficients (\%)}} \\
Individual: Actor $a_i^{(1)}$               & 17.95 & 18.37 & 0.25 & 2.34 & 1.79 & 0.80 \\
Individual: Partner $p_j^{(1)}$             & 16.67 & 17.30 & 0.27 & 2.22 & 1.92 & 0.77 \\
Individual: Dyad $d_{i,j}^{(1)}$            & 19.23 & 19.24 & 0.22 & 1.74 & 1.52 & 0.67 \\
Household: Actor $a_{h(i)}^{(2)}$           &  2.56 &  2.02 & 0.15 & 1.64 & 1.04 & 0.53 \\
Household: Partner $p_{h(j)}^{(2)}$         &  2.56 &  2.07 & 0.17 & 1.62 & 1.24 & 0.57 \\
Household: Dyad $d_{h(i),h(j)}^{(2)}$       & 41.03 & 40.99 & 0.38 & 3.14 & 2.70 & 1.12 \\
& & & & & & \\
\multicolumn{7}{l}{\emph{Reciprocity correlations}} \\
Individual: Generalised $\rho_{ap(1)}$      &  0.90 & 0.88 & 0.005 & 0.03 & 0.03 & 0.01 \\
Individual: Dyadic $\rho_{dd(1)}$           &  0.90 & 0.88 & 0.005 & 0.05 & 0.03 & 0.01 \\
Household: Generalised $\rho_{ap(2)}$       &  0.40 & 0.08 & 0.02 & 0.42 & 0.15 & 0.08 \\
Household: Dyadic $\rho_{dd(2)}$            &  0.90 & 0.89 & 0.002 & 0.02 & 0.02 & 0.01 \\
\hline 
\end{tabularx}
\end{footnotesize}
\end{table}

\newpage
\section{Extension of multilevel Social Relations Model to include within-household ties}

The structural model for $z_{i,j}$, the propensity of a $i \rightarrow j$ tie between individuals $i$ and $j$ in different households, given by eq. (\ref{eq:model_structural}) can be extended to incorporate within-household ties.  As before, an individual-level effect is denoted by superscript (1) and a household-level effect by (2) and $h(i)$ and $h(j)$ denote the households of individuals $i$ and $j$. In addition we distinguish individual, household and dyad effects for within-household and between-household ties by superscripts W and B respectively. A joint model for ties between individuals in the same household, $h(i)=h(j)$, and ties between individuals in different households, $h(i) \neq h(j)$, is specified as

\begin{eqnarray}
\label{eq:model_structural_bw}
z_{i,j} & = & \mbox{I} \left (h(i) = h(j) \right ) \left [ a_i^{(W1)} + p_j^{(W1)} + d_{i,j}^{(W1)} + v_{h(i)}^{(W2)} \right ] \\ \nonumber
 & + & \mbox{I} \left (h(i) \neq h(j) \right ) \left [ a_i^{(B1)} + p_j^{(B1)} + d_{i,j}^{(B1)} + a_{h(i)}^{(B2)} + p_{h(j)}^{(B2)} + d_{h(i),h(j)}^{(B2)} \right ]
\end{eqnarray}

where $\mbox{I}(\cdot)$ is the indicator function. The first line of eq. (\ref{eq:model_structural_bw}) specifies the sub-model for within-household ties and the second line, equivalent to eq. (\ref{eq:model_structural}), is the sub-model for between-household ties. It is not possible to distinguish actor and partner effects at the household level because the actor and partner are from the same household. For this reason, there is only one household-level random effect in the within model, $v_{h(i)}^{(W2)}$, which represents a household's tendency to engage in within-person exchanges.

The within and between components of eq. (\ref{eq:model_structural_bw}) are linked by allowing for correlation among the random effects at a given level.  We make the following distributional assumptions:

\begin{equation*}
\begin{gathered}
\begin{pmatrix}
a_i^{(W1)} \\
p_i^{(W1)} \\
a_i^{(B1)} \\
p_i^{(B1)} \\
\end{pmatrix}
\sim N \left ( \bm{0}, \Sigma_{ap(WB1)} \right ), \quad
\Sigma_{ap(WB1)} = \begin{pmatrix}
\sigma_{a(W1)}^2 & & & \\
\sigma_{ap{(W1)}} & \sigma_{p(W1)}^2 & & \\
\sigma_{aa{(WB1)}} & \sigma_{ap{(WB1)}} & \sigma_{a(B1)}^2 &  \\
\sigma_{pa{(WB1)}} & \sigma_{pp{(WB1)}} & \sigma_{pa{(B1)}} & \sigma_{p(B1)}^2   \\
\end{pmatrix} \\[2ex]
\begin{pmatrix}
d_{i,j}^{(W1)} \\
d_{j,i}^{(W1)} \\
\end{pmatrix}
\sim N \left ( \bm{0}, \Sigma_{d(W1)} \right ), \quad
\Sigma_{d(W1)} = \begin{pmatrix}
\sigma_{d(W1)}^2 & \\
\sigma_{dd(W1)} & \sigma_{d(W1)}^2  \\
\end{pmatrix} \\[2ex]
\begin{pmatrix}
d_{i,j}^{(B1)} \\
d_{j,i}^{(B1)} \\
\end{pmatrix}
\sim N \left ( \bm{0}, \Sigma_{d(B1)} \right ), \quad
\Sigma_{d(B1)} = \begin{pmatrix}
\sigma_{d(B1)}^2 & \\
\sigma_{dd(B1)} & \sigma_{d(B1)}^2  \\
\end{pmatrix} \\[2ex]
\begin{pmatrix}
v_{h(i)}^{(W2)} \\
a_{h(i)}^{(B2)} \\
p_{h(i)}^{(B2)} \\
\end{pmatrix}
\sim N \left ( \bm{0}, \Sigma_{ap(WB2)} \right ), \quad
\Sigma_{ap(WB2)} = \begin{pmatrix}
\sigma_{v(W2)}^2 & \\
\sigma_{av{(WB2)}} & \sigma_{a(B2)}^2 & \\
\sigma_{pv{(WB2)}} & \sigma_{pa{(B2)}} & \sigma_{p(B2)}^2 & \\
\end{pmatrix} \\[2ex]
\begin{pmatrix}
d_{h(i), h(j)}^{(B2)} \\
d_{h(j), h(i)}^{(B2)} \\
\end{pmatrix}
\sim N \left ( \bm{0}, \Sigma_{d(B2)} \right ), \quad
\Sigma_{d(B2)} = \begin{pmatrix}
\sigma_{d(B2)}^2 & \\
\sigma_{dd{(B2)}} & \sigma_{d(B2)}^2 \\
\end{pmatrix}.
\end{gathered}
\end{equation*}

The measurement model of eq. (\ref{eq:model_measurement}) can be extended to handle within-household ties in a similar way.  For within-household exchanges, with $h(i)=h(j)=h(m)$, we cannot distinguish actor and partner effects $r_{a,h(i)}^{(2)}$ and $r_{p,h(j)}^{(2)}$ at the household level.  While we might consider replacing the actor and partner effects by a single household-level random effect $r_{h(i)}^{(W2)}$, this would be confounded with $v_{h(i)}^{(W2)}$ in the structural model of eq. (\ref{eq:model_structural_bw}).  The reporter household dyad effect $r_{d|h(i)h(j)|h(m)}^{(2)}$ also reduces to an overall household effect when the actor, partner and reporter households coincide, which again cannot be separated from the household effect in the structural model.  Thus all household effects are excluded from the within-household component of the measurement model.

\section{Derivation of expression for the mean probability of a between-individual tie}

Denote by $y_{i,j,m}^*$ the continuous latent response underlying the observed binary indicator $y_{i,j,m}$ of a $i \rightarrow j$ tie from actor $i$ to partner $j$ reported by individual $m \in \{i,j\}$ such that $y_{i,j,m} = \mbox{I}(y_{i,j,m}^* > 0)$.   Combining the measurement and structural parts of the multilevel probit model, given by eq. (\ref{eq:model_measurement}) and (\ref{eq:model_structural}), the model can be expressed in terms of $y_{i,j,m}^*$ as
\begin{equation*}
y_{i,j,m}^* = \eta_{i,j} + w_{i,j,m}
\end{equation*}
where
\begin{equation*}
\eta_{i,j}= 0.5(\beta_a + \beta_p) + a_i^{(1)} + p_j^{(1)} + d_{i,j}^{(1)} + a_{h(i)}^{(2)} + p_{h(j)}^{(2)} + d_{h(i),h(j)}^{(2)}
\end{equation*}
and
\begin{equation*}
 w_{i,j,m} = \mbox{I}(m=i) (r_{ai}^{(1)} + r_{a,h(i)}^{(2)}) + \mbox{I}(m=j) (r_{pj}^{(1)} + r_{p,h(j)}^{(2)}) 
 + r_{d |i,j| m}^{(1)} + r_{d |h(i),h(j)| h(m)}^{(2)} + \epsilon_{i,j,m}.
\end{equation*}

To derive the household network from the model (described in Section 4.2 of the paper), we first obtain  the mean probability of a tie between actor individual $i$ and partner individual $j$. This is obtained by conditioning on the component of the model $\eta_{i,j}$ that involves the fixed part parameters and individual and household random effects, and integrating out the reporter random effects in $w_{i,j,m}$.

As $w_{i,j,m}$ is a linear combination of zero-mean independent and normally distributed random effects, $w_{i,j,m} \sim N(0, \sigma_w^2)$ where
\begin{equation*}
\sigma_w^2 = \mbox{I}(m=i) (\sigma_{ra(1)}^2 +\sigma_{ra(2)}^2) + \mbox{I}(m=j) (\sigma_{rp(1)}^2 +\sigma_{rp(2)}^2) + \sigma_{rd(1)}^2 + \sigma_{rd(2)}^2 + 1.
\end{equation*}

Let $w_{i,j,m}^* = w_{i,j,m}/\sigma_w \sim N(0,1)$ then, generalising \cite[][Appendix 1: Section 1.1]{bland.cook2019} from a two-level  random intercept logit model, the target probability is derived as

\begin{align*}
\pi_{i,j} & = \mathbb{P}(Y_{i,j,m}^* > 0 \, | \, \eta_{i,j}) \\
& = \mathbb{P}(\eta_{i,j} + w_{i,j,m} > 0 \, | \, \eta_{i,j}) \\
& = \mathbb{P}\left ( w_{i,j,m}^* > - \frac{\eta_{i,j}}{\sigma_w} \, | \, \eta_{i,j} \right ) \\
& = \mathbb{P}\left ( w_{i,j,m}^* < \frac{\eta_{i,j}}{\sigma_w} \, | \, \eta_{i,j} \right ) \\
& = \Phi \left ( \frac{\eta_{i,j}}{\sigma_w}  \right )
\end{align*}
where $\Phi(\cdot)$ is the standard normal cdf.

Both $\eta_{i,j}$ and $\sigma_w^2$ depend on whether the reporter $m$ is the actor or the partner in the $(i,j)$ dyad. As the sample observations are split equally between actor and partner reports, we can replace $\mbox{I}(m=i) \beta_a + \mbox{I}(m=j) \beta_p$ by the average intercept $0.5 (\beta_a + \beta_p)$ in $\eta_{i,j}$ and $\mbox{I}(m=i) (\sigma_{ra(1)}^2 +\sigma_{ra(2)}^2) + \mbox{I}(m=j)(\sigma_{rp(1)}^2 +\sigma_{rp(2)}^2)$ by the average variance $0.5(\sigma_{ra(1)}^2 +\sigma_{ra(2)}^2 + \sigma_{rp(1)}^2 +\sigma_{rp(2)}^2)$ in $\sigma_w^2$.  Alternatively, we can compute predicted probabilities of $\pi_{i,j}$ for $m=i$ and $m=j$ and take their average.

\newpage

\section{Posterior predictive model checks}

The fit of the final model was assessed using posterior predictive checks based on replicates of the binary reports of ties $y_{i,j,m}$ simulated from the MCMC parameter chains for the fitted model, including predicted values of all random effects. Test statistics were chosen to assess the model's ability to capture several features of the data: (i) within-dyad correlations, and (ii) standard deviations of out-degree and in-degree across individuals and households, separately for actor and partner reports. For the household statistics, the household network was constructed using the union and density methods of aggregation: the union network is based on a binary indicator of any support given by an individual in household $k$ to an individual in household $l$, while the density network is based on the sample proportion of individual support ties in a $k \rightarrow l$ household dyad. Table \ref{tab:post_pred_checks} shows the mean and 2.5 and 97.5 percentiles of the distribution of each test statistic across replicates. In each case, the observed value of the test statistic is close to the posterior predictive mean and lies well within the 2.5 and 97.5 percentiles of the posterior predictive distribution, indicating that the model provides a good fit to these aspects of the data.

\begin{table}[!htbp]
\centering
\caption{Posterior predictive checks for the final multilevel SRM, comparing statistics calculated from the observed data $y_{i,j,m}$ with their posterior distributions based on replicates simulated from the fitted model}
\label{tab:post_pred_checks}
\begin{footnotesize}
\begin{tabularx}{0.75\textwidth}{l *{4}{R}}
\hline
& & \multicolumn{3}{c}{Posterior predictive distribution} \\
Statistic & Observed & Mean & 2.5\%  & 97.5\% \\
\hline
Within-dyad correlations & & & & \\
~~~$\mbox{cor}(y_{i,j,i},y_{i,j,j})$ & 0.312 & 0.315 & 0.295 & 0.336 \\
~~~$\mbox{cor}(y_{i,j,i},y_{j,i,i})$ & 0.606 & 0.603 & 0.582 & 0.623 \\
~~~$\mbox{cor}(y_{i,j,i},y_{j,i,j})$ & 0.351 & 0.337 & 0.311 & 0.364 \\
~~~$\mbox{cor}(y_{i,j,j},y_{j,i,i})$ & 0.248 & 0.255 & 0.228 & 0.282 \\
& & & & \\
SD of individual out- and in-degree & & & & \\
~~~Out-degree: actor reports & 16.27 & 16.32 & 15.70 & 16.95 \\
~~~Out-degree: partner reports & 12.84 & 12.76 & 12.16 & 13.36 \\
~~~In-degree: actor reports & 14.35 & 14.17 & 13.53 & 14.81 \\
~~~In-degree: partner reports & 16.08 & 16.18 & 15.53 & 16.83 \\
& & & & \\
SD of household out- and in-degree & & & & \\
\emph{Union network} & & & & \\
~~~Out-degree: actor reports & 6.44 & 6.22 & 5.67 & 6.75 \\
~~~Out-degree: partner reports & 4.42 & 4.49 & 3.92 & 5.05 \\
~~~In-degree: actor reports & 5.21 & 4.86 & 4.29 & 5.42 \\
~~~In-degree: partner reports & 7.58 & 7.33 & 6.80 & 7.85 \\
\emph{Density network} & & & & \\
~~~Out-degree: actor reports & 3.33 & 3.30 & 3.06 & 3.54 \\
~~~Out-degree: partner reports & 2.76 & 2.73 & 2.51 & 2.95 \\
~~~In-degree: actor reports & 2.88 & 2.87 & 2.65 & 3.09 \\
~~~In-degree: partner reports & 3.31 & 3.34 & 3.09 & 3.59 \\
\hline
\end{tabularx}
\end{footnotesize}
\end{table}

\newpage
\section{Further analysis of alternative household networks for social support in Nicaragua}

\subsection{Plots of household centrality estimates by network type}

\begin{figure} [htbp]
    \centering
    \includegraphics[width=0.7\linewidth]{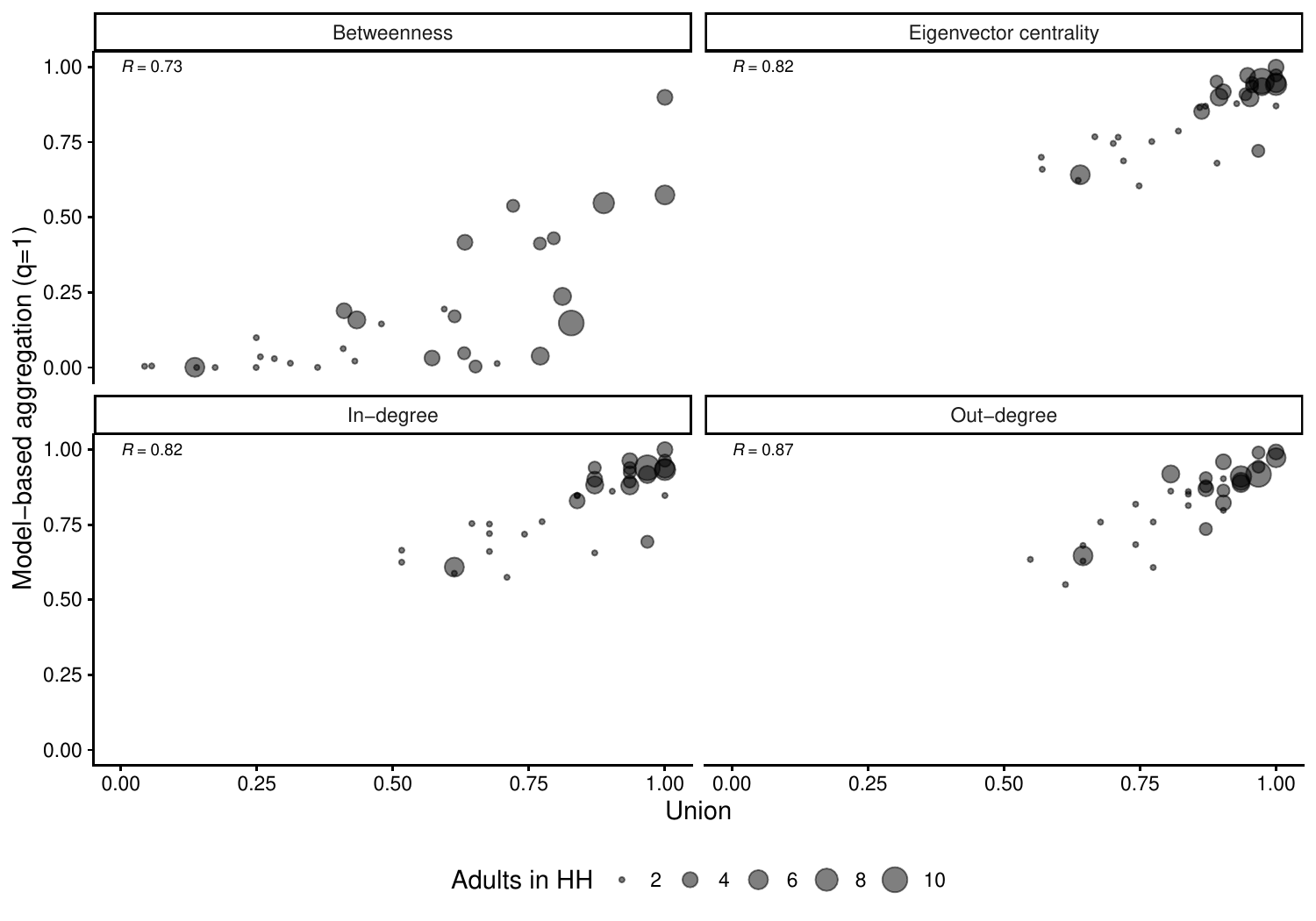}
    \caption{Household centrality estimates for aggregation networks: the union ($y_{k,l}^u$) and model-based ($\hat{\theta}^{**}_{k,l}(1)$) networks. Estimates are normalised by dividing each measure by its maximum.}
    \label{fig:agg_cen}
\end{figure}

\begin{figure} [htbp]
    \centering
    \includegraphics[width=0.7\linewidth]{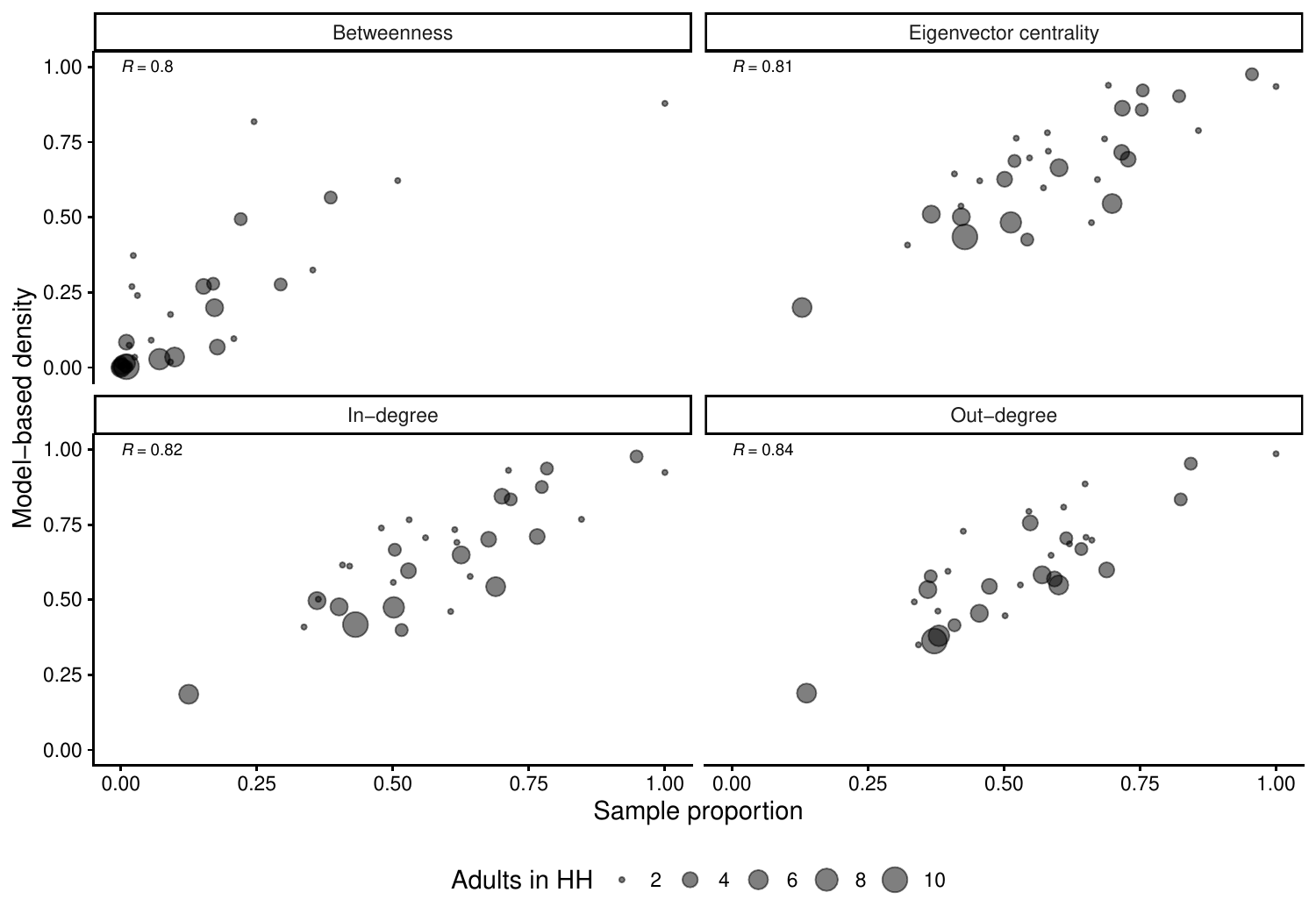}
    \caption{Household centrality estimates for density networks: the sample proportion ($p_{k,l}$) and model-based ($\hat{\theta}^{*}_{k,l}$) networks. Estimates are normalised by dividing each measure by its maximum.}
    \label{fig:dens_cen}
\end{figure}

\begin{figure} [htbp]
    \centering
    \includegraphics[width=0.7\linewidth]{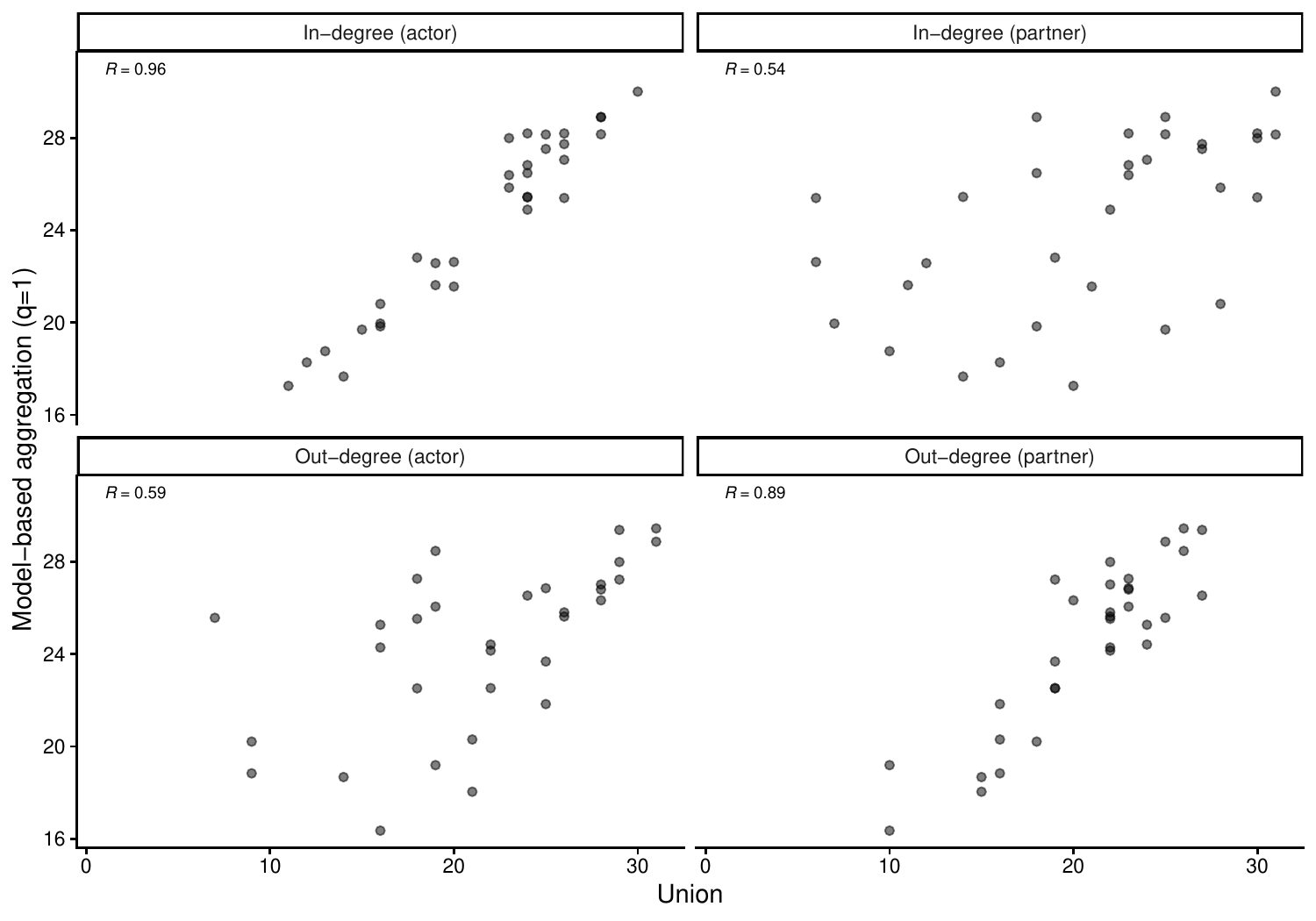}
    \caption{Household degree centrality estimates for aggregation networks: the union networks ($y_{k,l}^{u(a)}$ and $y_{k,l}^{u(p)}$) use actor or partner reports while the model-based network ($\hat{\theta}^{**}_{k,l}(1)$) uses combined reports.}
    \label{fig:app_ap_deg}
\end{figure}

\begin{figure} [htbp]
    \centering
    \includegraphics[width=0.7\linewidth]{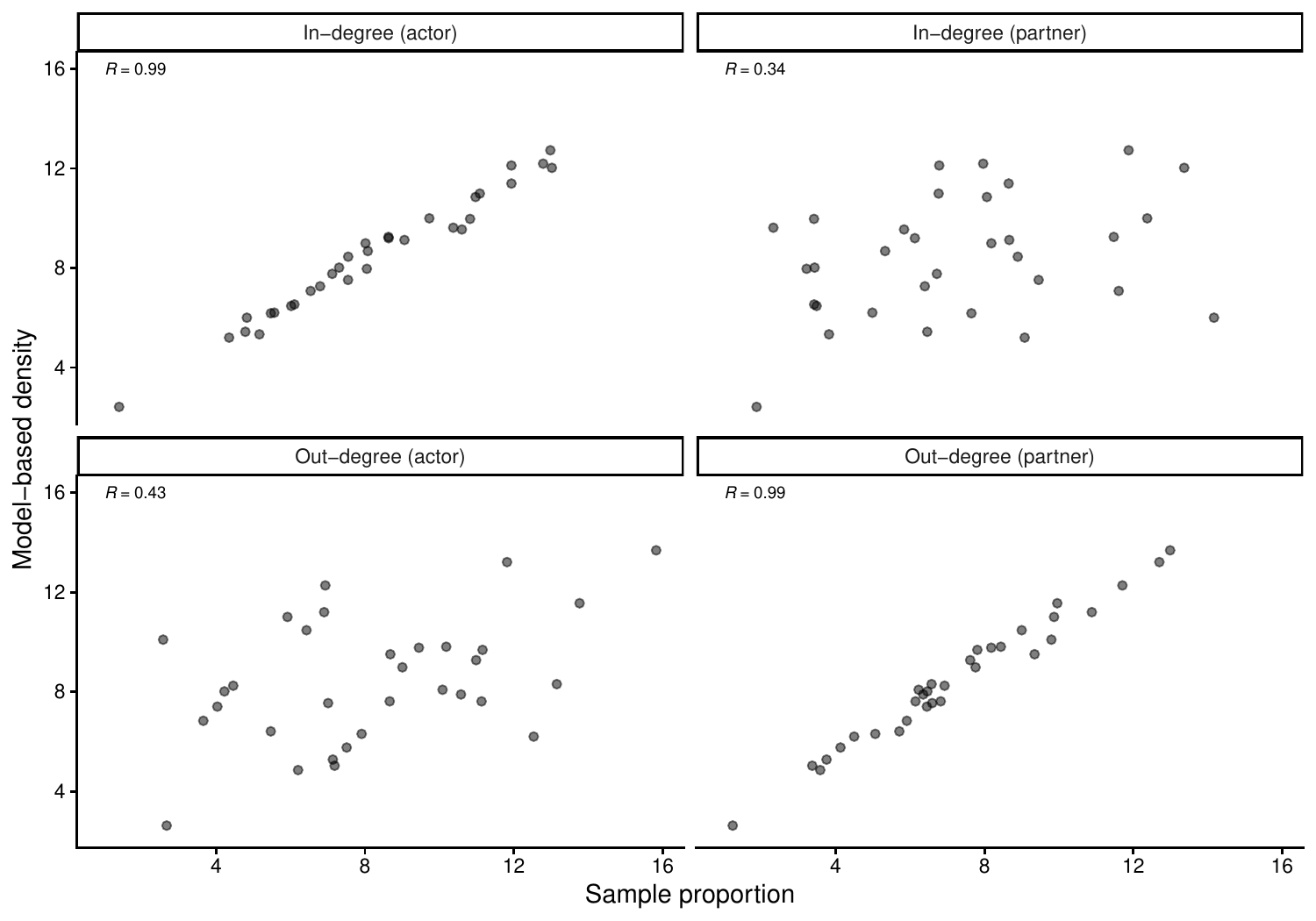}
    \caption{Household degree centrality estimates for density networks: the sample proportion networks ($p_{k,l}^{(a)}$ and $p_{k,l}^{(p)}$) use actor or partner reports while the model-based network ($\hat{\theta}^{*}_{k,l}$) uses combined reports.}
    \label{fig:dens_ap_deg}
\end{figure}

\begin{figure}
    \centering
    \includegraphics[width=0.7\linewidth]{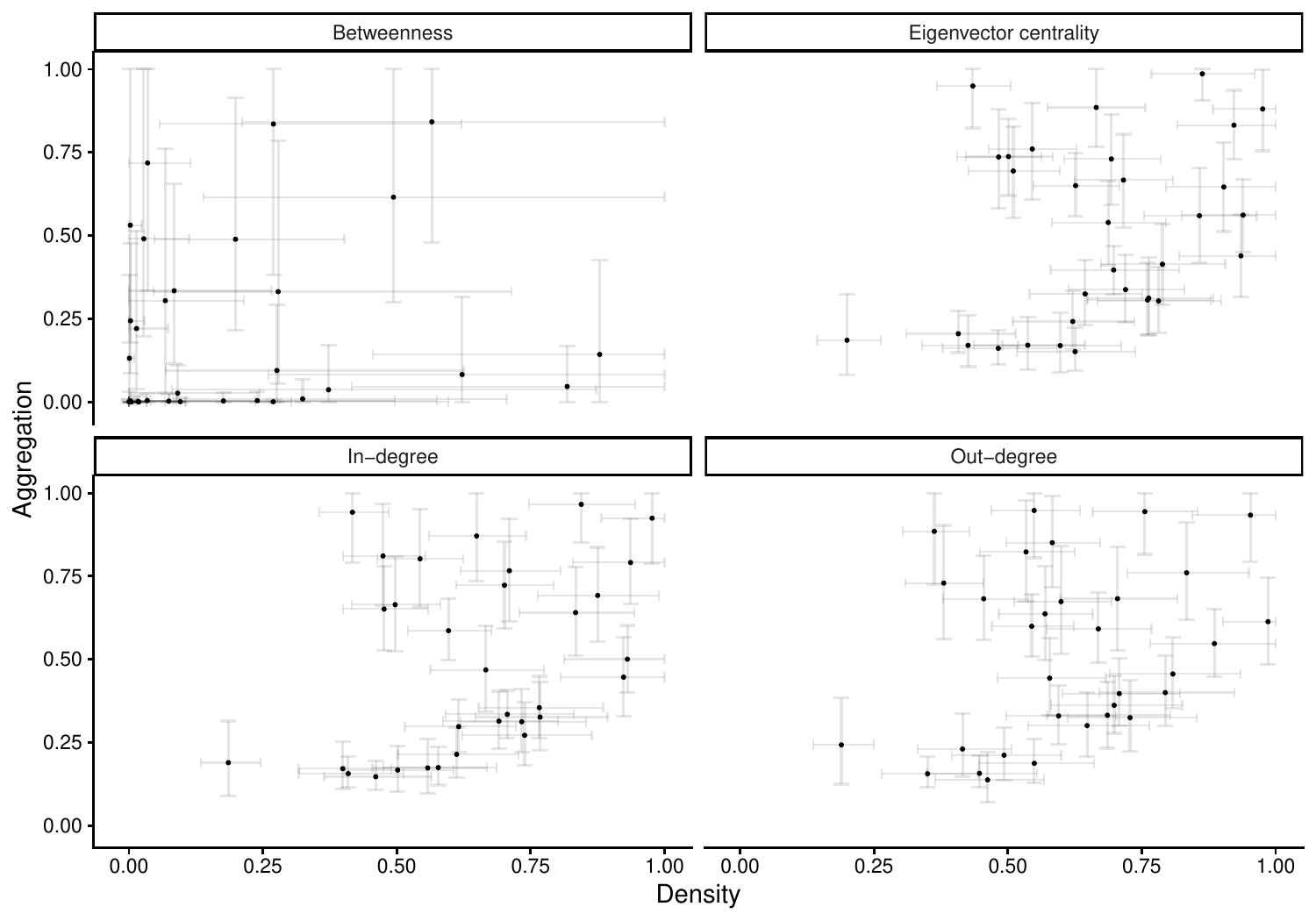}
    \caption{Household centrality estimates for model-based aggregation ($q=4$, $\hat{\theta}^{**}_{k,l}(4)$) and density ($\hat{\theta}^{*}_{k,l}$) methods with 95\% credible intervals. Estimates are normalised by dividing each measure by its maximum.}
    \label{fig:hh_centrality_credible_intervals}
\end{figure}

\clearpage

\subsection{Impact of shrinkage in random effect estimates on predicted household tie probabilities}
As noted at the end of Section 4.2 of the paper, one advantage of constructing a household network from the multilevel Social Relations Model is that model-based predictions of between-household ties are ``precision-weighted'' because they are calculated from empirical Bayes (EB) estimates of the random effects which are ``shrunken'' towards their means of zero when based on small sample sizes.  In this section, we consider whether discrepancies between estimates of the predicted probabilities of household ties $\hat{\theta}^{*}_{k,l}$ (i.e. the model-based density network) and the sample proportions $p_{k,l}$ are consistent with shrinkage.  

It is important to note that in a complex multilevel model with multiple random effects shrinkage does not necessarily pull $\hat{\theta}^{*}_{k,l}$ towards its overall mean. Shrinkage can shift $\hat{\theta}^{*}_{k,l}$ away from the mean because some random effects are more subject to shrinkage than others and random effects vary in the strength of their contribution to $\hat{\theta}^{*}_{k,l}$. As the calculation of $\hat{\theta}^{*}_{k,l}$ involves integrating out all reporter effects in the measurement model and averaging over individual effects in the structural model, the amount of shrinkage in $\hat{\theta}^{*}_{k,l}$ is determined primarily by the EB estimates of the household random effects $(a_k^{(2)}, \, p_k^{(2)}, \, d_{k,l}^{(2)})$.  Among these household effects, estimates of the household dyad effects $d_{k,l}^{(2)}$ have the largest shrinkage because, under double-sampling, they are based on $2 n_k n_l$ observations (ranging from 6 to 140 for the Nicaragua data), while the household actor and partner effects $(a_k^{(2)}, \, p_k^{(2)}$ are based on much larger sample sizes (316 to 1940 observations). Moreover, as $d_{k,l}^{(2)}$ accounts for 41.9\% of the total variance in $z_{i,j}$ (see Table 3 in the paper), shrinkage in its EB estimate can have a substantial impact on $\hat{\theta}_{k,l}^*$.  If $p_{k,l}$ is large, $d_{k,l}^{(2)}$ will be large and positive, so shrinkage will pull its EB estimate towards its mean of zero which will pull $\hat{\theta}_{k,l}^*$ downwards.  Similarly, if $p_{k,l}$ is small, $d_{k,l}^{(2)}$ will be large and negative, so shrinkage will pull $\hat{\theta}_{k,l}^*$ upwards.

\begin{figure} [htbp]
    \centering
    \includegraphics[width=0.7\linewidth]{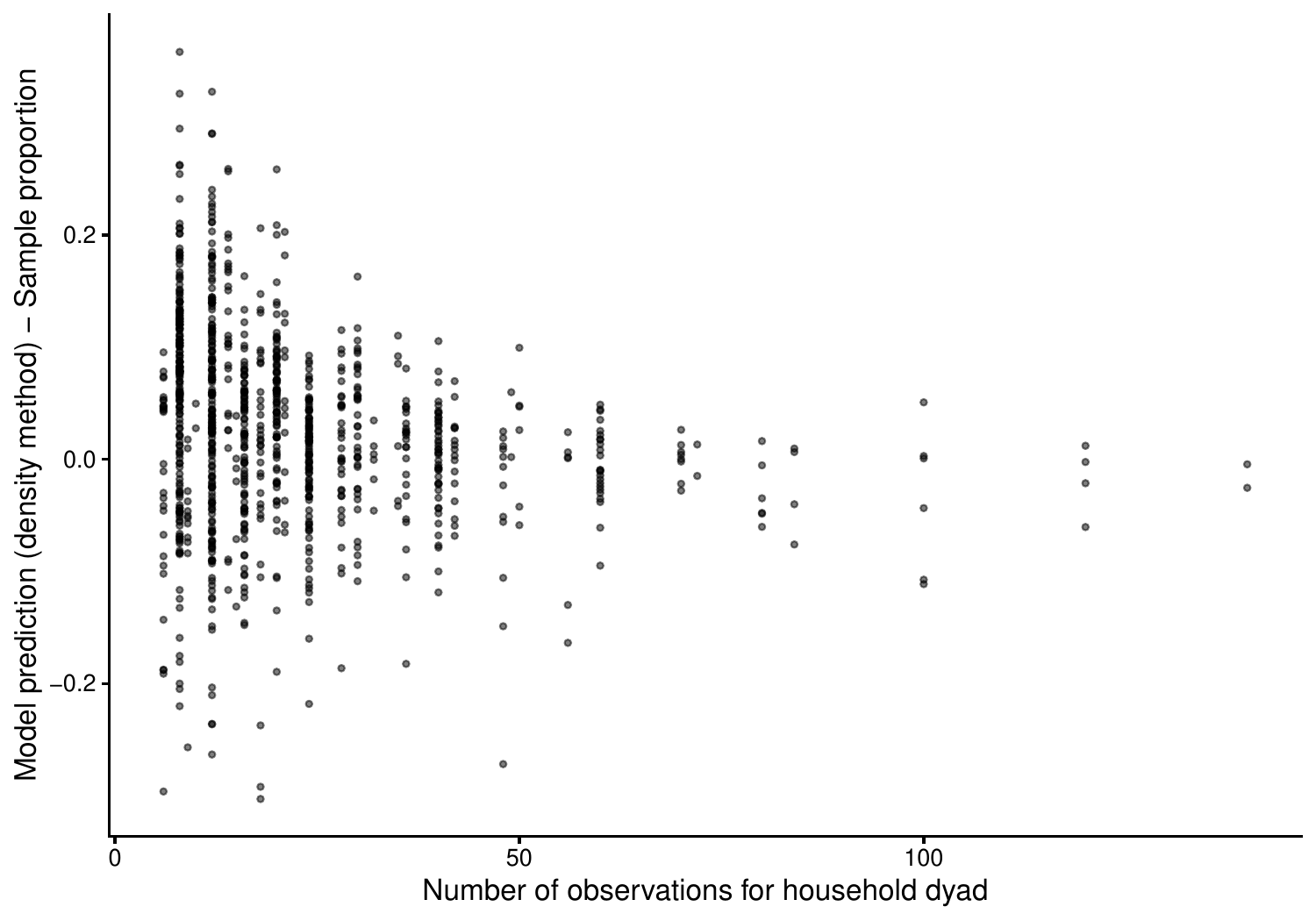}
    \caption{Distribution of difference between model-based estimates of between-household tie probabilities ($\hat{\theta}_{k,l}^*$) and sample proportions ($p_{k,l}$) by number of observations in the household dyad ($2 n_k n_l$).}
    \label{fig:dens_shrink}
\end{figure}

Figure \ref{fig:dens_shrink} plots $\hat{\theta}_{k,l}^* - p_{k,l}$ against $2 n_k n_l$.  We find larger discrepancies between $\hat{\theta}_{k,l}^*$ and $p_{k,l}$ when $2 n_k n_l$ is small, which is consistent with greater shrinkage in the EB estimates of $d_{k,l}^{(2)}$ when based on fewer observations. 
To further explore the impact of shrinkage on $\hat{\theta}_{k,l}^*$, we can compare the distributions of $\hat{\theta}_{k,l}^*$ and $p_{k,l}$. The distribution of $p_{k,l}$ across household dyads is asymmetric with a high proportion of zeros (16.7\%) and few ones (1.7\%).  Among dyads with a below-average $p_{k,l}$ the correlation between $\hat{\theta}_{k,l}^* - p_{k,l}$ and $2 n_k n_l$ is $-$0.42, which is consistent with the expected impact of shrinkage described above: smaller $2 n_k n_l$ is associated with a larger (and positive) $\hat{\theta}_{k,l}^* - p_{k,l}$.  Shrinkage also leads to a reduction in the standard deviation of probabilities across household dyads: 0.36 for $p_{k,l}$ and 0.21 for $\hat{\theta}_{k,l}^*$. 

We might expect that the dependence of $\hat{\theta}_{k,l}^* - p_{k,l}$ on the number of observations for household $(k,l)$ would translate to a dependence of the difference in their respective household centrality estimates on household sample size (because households with small $n_k$ will have smaller values for $n_k n_l$). However, the discrepancy between household centrality estimates based on $\hat{\theta}_{k,l}^*$ and $p_{k,l}$ do not appear to be related to the number of adults in the household for any centrality measure (see Figure \ref{fig:dens_cen}). This is because a model-based centrality estimate for household $k$ takes into account \emph{all} $\hat{\theta}_{k,l}^*$ for $l \neq k$ and $\hat{\theta}_{k,l}^* - p_{k,l}$ depends on shrinkage in the EB estimate of $d_{k,l}^{(2)}$ which varies across partner households $l$, so that $\hat{\theta}_{k,l}^* - p_{k,l}$ can vary in magnitude and direction with $l$.  Consider, for example, a comparison between the model-based density estimate of out-degree for household $k$, calculated as  $\sum_{l \neq k} \hat{\theta}_{k,l}^*$, and the corresponding measure based on sample proportions, $\sum_{l \neq k} p_{k,l}$.  As noted above, shrinkage in the estimate of $d_{k,l}^{(2)}$ may pull $\hat{\theta}_{k,l}^*$ above or below $p_{k,l}$, thus $\sum_{l \neq k} \hat{\theta}_{k,l}^*$ may be higher or lower than $\sum_{l \neq k} p_{k,l}$ and the magnitude of the difference will not necessarily be related to $n_k$.

\newpage

\subsection{An example of outliers in centrality measures from simple aggregations}

As outlined in the main text and the preceding section, the model-based approaches to estimating household centrality measures offer several advantages over estimates based on simple descriptive methods for constructing household networks. In comparison to prevailing methods that rely on aggregation and dichotomisation, the model-based approaches make fuller use of the available information.

\begin{figure} [htbp]
    \centering
    \includegraphics[width=0.8\linewidth]{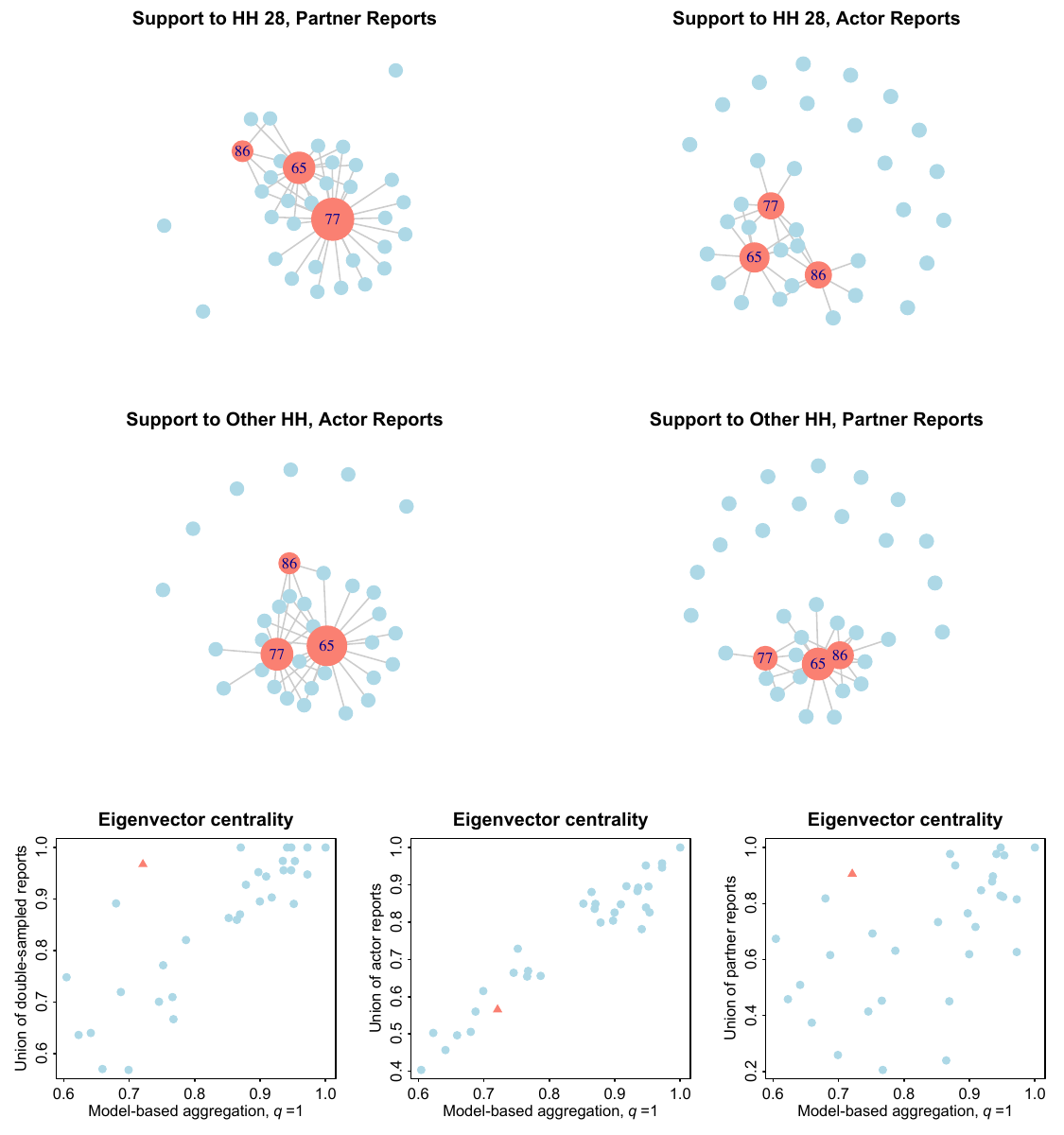} 
    \caption{The network visualisations depict ties between individuals in Household 28, as represented by the numbered nodes, and other households in the community. The data have been dichotomised such that a tie appears if there is a nomination between an individual in Household 28 and any individual in the other household. The two visualisations on the left represent nominations made by the individuals in Household 28 for the respective double-sampled questions. The visualisations on the right depict nominations by members of other households.
    These data are further dichotomised to create three binary household-to-household networks based on the union of the double-sampled data ($y_{k,l}^u$), the union of the actor reports ($y_{k,l}^{u(a)}$), and the union of the partner reports ($y_{k,l}^{u(p)}$). For these three networks, estimates of the households' eigenvector centrality are compared to the corresponding estimate from the model-based aggregation network ($\hat{\theta}^{**}_{k,l}(1)$). In these scatterplots, the salmon-coloured triangles represent Household 28.
    }
    \label{fig:HH28}
\end{figure}

The Nicaragua data provide an opportunity to illustrate how substantially the model-based centrality estimates can differ from those based on descriptive statistics. Consider the case of Household 28, which includes three adults: a 65-year old male household head (ID code 86), his 49-year-old wife (ID code 65), and the wife's 33-year-old nephew (ID code 77). For both of the double-sampled questions (on giving and receiving support), Figure \ref{fig:HH28} depicts the reported ties involving these individuals and other households in the community. When these individuals are in the role of reporters, they collectively report at least one connection to most other households -- 90\% in the case of actor reports (of giving support) and 81\% for the partner reports (of receiving support). But conversely, for both of the double-sampled questions, only 52\% of the other 31 households have reporters who nominated individuals from Household 28.

Figure \ref{fig:HH28} also shows the household eigenvector centrality estimates from four constructions of the household-to-household network: (1) the model-based aggregation network, where \textit{q}=1, (2) the union of the double-sampled reports, (3) the union of the actor reports, and (4) the union of the partner reports. In the case of Household 28, there are noteworthy differences among the eigenvector centrality estimates for these network constructions. That is, Household 28 is in the lowest 25th percentile for the model-based aggregation network and the union of actor reports. By contrast, the household is in the highest 25th percentile for the union of the double-sampled reports and the union of the partner reports.

The divergent estimates for Household 28 underscore the extent to which centrality measures can be sensitive to the use of binary vs weighted networks and to measurement considerations. It is beneficial to have datasets with multiple reported observations of the same dyads, and it is beneficial to employ statistical approaches that can leverage the information in those datasets.

\end{document}